\documentclass[prd,preprint,tightenlines,floatfix,showpacs,preprintnumbers,nofootinbib]{revtex4}
 \usepackage[dvips,final]{graphicx}
  \usepackage{amssymb,amsmath,epsfig,bm,pifont}
   \usepackage{pstricks,pst-node,pst-text,pst-3d}
\usepackage{color}
 
  \newcommand{\BluTn}[1]{\textcolor{blue}{#1}}
   \newcommand{\RedTn}[1]{\textcolor{red}{#1}}
    \newrgbcolor{DarkGreen}{0.04 0.5 0.1}
\newcommand{\Ds}{\displaystyle}                           
\def\1{\hbox{{1}\kern-.25em\hbox{l}}}

\begin{document}
\thispagestyle{empty}
\date{\today}
\preprint{\hbox{RUB-TPII-03/2010}}

\title{Higher-order QCD perturbation theory in different schemes:
       From FOPT to CIPT to FAPT\\}
      \author{A.~P.~Bakulev}
\email{bakulev@theor.jinr.ru}
\affiliation{Bogoliubov Laboratory of Theoretical Physics, JINR,
             141980 Dubna, Russia\\}

\author{S.~V.~Mikhailov}%
\email{mikhs@theor.jinr.ru}
\affiliation{Bogoliubov Laboratory of Theoretical Physics, JINR,
             141980 Dubna, Russia\\}

\author{N.~G.~Stefanis}
\email{stefanis@tp2.ruhr-uni-bochum.de}
\affiliation{Institut f\"{u}r Theoretische Physik II,
             Ruhr-Universit\"{a}t Bochum,
             D-44780 Bochum, Germany\\}
\affiliation{Bogoliubov Laboratory of Theoretical Physics, JINR,
             141980 Dubna, Russia\\}

\vspace {10mm}
\begin{abstract}
Results on the resummation of non-power-series expansions of the
Adler function of a scalar, $D_S$, and a vector, $D_V$, correlator
are presented within fractional analytic perturbation theory (FAPT).
The first observable can be used to determine the decay width
$\Gamma_{H\to b\bar{b}}$ of a scalar Higgs boson to a
bottom-antibottom pair, while the second one is relevant for
the $e^+e^-$ annihilation cross section.
The obtained estimates are compared with those from fixed-order
(FOPT) and contour-improved perturbation theory (CIPT),
working out the differences.
We prove that although FAPT and CIPT are conceptually different,
they yield identical results.
The convergence properties of these expansions are discussed
and predictions are extracted for the resummed series of $R_S$
and $D_V$ using one- and two-loop coupling running,
and making use of appropriate generating functions
for the coefficients of the perturbative series.
\end{abstract}
\maketitle


\section{Introduction}
\label{sec:intro}
Since the original work of Shirkov and Solovtsov \cite{SS97} appeared
in 1997, the analytic approach to QCD perturbation theory has
evolved and progressed considerably.
At the heart of this approach is the spectral density which provides
the means to define an analytic running coupling in the Euclidean
space via a dispersion relation in accordance with causality and
renormalization-group (RG) invariance.
Using the same spectral density one can also define the running
coupling in Minkowski space by employing the dispersion relation for
the Adler function \cite{Rad82,KP82,MS96,MiSol97}.
In parallel, this approach has been extended beyond the one-loop level
\cite{MiSol97,SS98} and analytic and numerical tools for its
application have been developed
\cite{Mag99,Mag00,KM01,Mag03u,KM03,Mag05}.
These efforts culminated in a systematic calculational framework,
termed Analytic Perturbation Theory (APT), recently reviewed in
\cite{SS06}.

The simple analytization of the running coupling and its integer
powers has been generalized to the level of hadronic amplitudes
\cite{KS01,Ste02} as a whole and new techniques have been
developed to deal with more than one (large) momentum scale
\cite{SSK99,SSK00,BPSS04} (for a brief exposition,
see \cite{Ste04}).
This encompassing version of analytization includes all terms
that may contribute to the spectral density, i.e., affect the
discontinuity across the cut along the negative real axis
$-\infty<Q^2<0$.
It turns out that logarithms of the second large scale, which can be
the factorization scale in higher-order perturbative calculations or
the evolution scale, correspond to non-integer indices of the
analytic couplings, giving rise---in Euclidean space---to
Fractional Analytic Perturbation Theory (FAPT) \cite{BMS05,BKS05}.
This concept was successfully extended to the timelike region and a
unified description in the whole complex momentum space was achieved
\cite{BMS06} (see also \cite{SK08,Ste09}).
The issue of crossing heavy-flavor thresholds, naturally entering such
calculations, has been considered within APT
\cite{Shi00,Shi01,SS06,BPSS04} and more recently also within FAPT
\cite{AB08gfapt}.

Another important topic, which is at the core of the present
investigation, deals with the perturbative summation in (F)APT.
As it has been demonstrated in \cite{MS04,BM08,CvVa06}, it is possible
to determine the total sum of the perturbative expansion at the
level of the one-loop running of the coupling.
This result will be extended here to include in the sum at least the
two-loop running of the coupling.
Even more, making a natural assumption concerning the asymptotic
behavior of the perturbative coefficients, proposed long ago by
Lipatov \cite{Lip76}, we will show that it is possible to estimate
 the sum of the series to all orders of the expansion.
This important feature of the (F)APT non-power series gives us the
possibility to estimate the uncertainties of the perturbative results
with higher accuracy relative to the conventional QCD power-series
expansion.
In the present investigation we will extend and systematize these
issues towards a complete calculational scheme considering also some
applications---including updated predictions for the decay width of a
Higgs boson to a bottom--antibottom pair, relevant for the Higgs-boson
search at the Tevatron and the LHC.

The paper is organized as follows.
In Sec.\ \ref{sec:APT.Elements}
we compare the results obtained in different
schemes---FOPT, CIPT, and FAPT---and prove that CIPT and FAPT
provide identical results for
$R_{e^+e^-\to\text{hadrons}}$ and $R_\text{S}$.
In addition, we recall in this section the pivotal ingredients of the
analytic perturbative approach and discuss briefly its extension
to the case of a variable flavor number
(the so-called ``global case'' \cite{Shi00}).
Moreover, in Sec.\ \ref{sec:APT.Resummation} we describe how typical
series expansions within FAPT can be resummed at the level of the
one-loop coupling running having recourse to a generating
function---referring the reader for the two-loop case
to Appendix~\ref{app:Two-loop}.
Section \ref{sec:App.Higgs} is devoted to the Higgs-boson decay into
a $\bar{b}b$ pair, a process which contains and exhibits the
conceptual and technical details of our analytic framework.
In the same section, we calculate the associated decay width,
which involves the Adler function of a scalar correlator, and
compare it with the results of the standard perturbative scheme.
Then, in Sec.\ \ref{sec:App.Adler}, we turn to another application of
the proposed resummation technique and consider the Adler function of
a vector correlator, pertaining to
$R_{e^+e^- \to \mbox{\footnotesize hadrons}}$.
Finally, Sec.\ \ref{sec:conclusions} contains our concluding remarks,
while some important technical derivations are given in five
Appendices.

\section{FOPT, CIPT and FAPT}
\label{sec:APT.Elements}
\subsection{Two-point correlator of scalar/vector currents}
\label{subsec:2-point.Correlator}
As mentioned in the Introduction, the initial motivation to invent
new QCD couplings was the desire to interrelate the Adler
$D$-function,
\begin{equation}
  D(Q^2,\mu^2) \stackrel{\mu^2=Q^2}{\longrightarrow} D(Q^2)
=
  d_0+ \sum_{m=1} d_m~a_s^m(Q^2)\, ,
\label{eq:D.Q}
\end{equation}
calculable in the Euclidean domain, and the quantity
$
 \Ds R_{e^+e^-}
=
 \frac{\sigma(e^+e^-
 \to \mbox{hadrons})}{\sigma(e^+e^- \to \mu^+\mu^-)}
$,
\begin{equation}
  R(s,\mu^2) \stackrel{\mu^2=s}{\longrightarrow} R(s)
=
  r_0 + \sum_{m=1} ~r_m~a_s^m(s)\, ,
\label{eq:R.s}
\end{equation}
which is measured in the Minkowski region.
Both quantities are considered in standard QCD perturbation theory,
demanding that the couplings entering them satisfy the
renormalization-group equation.
To facilitate direct comparison of our results further below
with the higher-order calculations in standard perturbation theory,
in particular Refs.\ \cite{BCK05,BCK08,BCK10}, we use here the variable
$a_s\equiv \alpha_s/\pi$.
Thus, the beta-function coefficients of this coupling
are defined by
\begin{equation}
 \label{eq:beta.s}
  \beta_s(a_s)
   =
  \frac{d\,a_s(\mu^2)}{d \ln(\mu^2)}
   =
  -a_s \left(\beta_{0}\ a_s
  +\beta_{1}\,a_s^2
  +\ldots
  \right)
  =
  -\frac{\alpha_s}{\pi}
    \left[b_0\,\left(\frac{\alpha_s}{4\,\pi}\right)
        + b_1\,\left(\frac{\alpha_s}{4\,\pi}\right)^2
        + \ldots
    \right]\,,
\end{equation}
where $\beta_n=b_n/4^{n+1}$ and the coefficients $b_n$
are specified in Appendix \ref{app:Two-loop}.

The functions $D$ and $R$ can be related to each other via the
following dispersion relations without any reference to
perturbation theory
\begin{equation}
  R(s)
=
  \hat{R}\left[D\right]
\equiv
  \frac{1}{2\pi \textit{i}}
  \int_{-s-i\varepsilon}^{-s+i\varepsilon}\!
  \frac{D(\sigma)}{\sigma}\,
  d\sigma\, , ~~~~~
\\
  D(Q^2)
=
  \hat{D}\left[R\right]
\equiv Q^2~\int_0^{\infty}
  \frac{R(\sigma)}{(\sigma+Q^2)^2}
  d\sigma~,
\label{eq:R-D-operation}
\end{equation}
where for the first term in Eq.\ (\ref{eq:R-D-operation}), the
integration contour $\Gamma_1$ around the cut (solid red line) is
shown in Fig.\ \ref{fig:G1G2G3}.
However, employing a perturbative expansion on the LHS of
Eqs.\ (\ref{eq:D.Q}) and (\ref{eq:R.s}), one obtains, in fact,
a relation between the powers of $\ln(s/\mu^2)$ and $\ln(Q^2/\mu^2)$
in the coefficients $r_m(s,\mu^2)$ and $d_n(Q^2,\mu^2)$, while the
powers of the couplings $a_s(\mu^2)$ reveal themselves as
parameters.

\subsection{Fixed-order perturbation theory}
\label{subsec:FOPT}
\begin{table}[b!]
\caption{This table exemplifies the structure of the first few
 coefficients $r_m=T^{mk}d_k$ (summation over $k=1,\ldots,m$ implied)
 of the conventional expansion of the $R$-ratio, where $T^{mk}$ are the
 table entries.
 Each coefficient $r_m$ contains a number of $d_{k}~(k\leq m)$
 terms in its expansion that are shown in the corresponding row.
 The various contributions are marked by different colors
 according to their loop order: one-loop---black; two-loop---red;
 three-loop---blue.
\label{Tab:r_n.d_k}}
\begin{tabular}{|c||p{24mm}|p{24mm}|p{24mm}|p{24mm}|p{24mm}|}\hline \hline
                     &\centerline{$\bm{d_1}$}
                                   &\centerline{$\bm{d_2}\vphantom{^\Big|}$}
                                                &\centerline{$\bm{d_3}$}
                                                             &\centerline{$\bm{d_4}$}
                                                                          &\centerline{$\bm{d_5}$}
\\ \hline \hline
          $\bm{r_1}\vphantom{^{\Big|}}$
                     & \centerline{$\bm{1}$}
                                   &            &            &            &
\\ \hline
          $\bm{r_2}\vphantom{^{\Big|}}$
                     & \centerline{$0$}
                                   & \centerline{$\bm{1}$}
                                                &            &            &
\\ \hline
          $\bm{r_3}\vphantom{^{\Big|}}$
                     & \centerline{$\Ds-\frac{(\pi\,\beta_0\vphantom{^{\big|}})^2}{3}\vphantom{^{\big|}_{\big|}}$}
                                   & \centerline{$0$}
                                                & \centerline{$\bm{1}$}
                                                             &            &
\\ \hline
          $\bm{r_4}\vphantom{^{\Big|}}$
                     &\centerline{$0$}
                      \centerline{\RedTn{$\Ds\bm{-\frac{5\,\pi^2\vphantom{^{\big|}}}{6}\,\beta_0\,\beta_1}\vphantom{^{\big|}_{\big|}}$}}
                                   & \centerline{$\Ds-\frac{(\pi \beta_0\vphantom{^{\big|}})^2}{3}~3\vphantom{^{\big|}_{\big|}}$}
                                                & \centerline{0}
                                                             &\centerline{$\bm{1}$}
                                                                          &
\\ \hline
          $\bm{r_5}\vphantom{^{\Big|}}$\footnote{
          \footnotesize This expression for $r_5$ was
          also obtained in Refs.\ \cite{KaSt95,DDGHMZ08}.}
                     &\centerline{$\Ds\frac{(\pi\,\beta_0\vphantom{^{\big|}})^4}{5}\vphantom{^{\big|}}$}\
                      \centerline{\RedTn{\bm{$\Ds -\frac{\pi^2}{2}\,\beta_1^2}$}}
                      \centerline{\BluTn{$\Ds \bm{-\pi^2\,\beta_0\,\beta_2}\vphantom{^{\big|}_{\big|}}$}}
                                   &\centerline{$\Ds0\vphantom{\frac{(\pi\,\beta_0^{\big|})^4}{5}_{\big|}}$}
                                    \centerline{\RedTn{$\Ds \bm{-\frac{7\,\pi^2}{3}\,\beta_0\,\beta_1}\vphantom{^{\big|}}$}}
                                                &\centerline{$\Ds-\frac{(\pi \beta_0)^2}{3}\,6$}
                                                             &\centerline{0}
                                                                          &\centerline{$\bm{1}$}
\\  \hline\hline
\end{tabular}
\end{table}
For the fixed order-$n$ perturbation theory (abbreviated as FOPT),
one can start from Eq.\ (\ref{eq:R-D-operation})
and use (\ref{eq:R.s})
on its LHS and
(\ref{eq:D.Q}) on its RHS
to obtain
\begin{eqnarray}
   R^{\text{FOPT}}_{n}(s)\equiv   R_{n}(s,\mu^{2}=s)
=
  \frac{1}{2\pi \textit{i}}
  \int_{-s-i\varepsilon}^{-s+i\varepsilon}\!
  \frac{D_{n}(\sigma,\mu^2)}
  {\sigma}\,
  d\sigma~
  \Bigg|_{\mu^2=s} = r_0+ \sum_{m=1}^n r_m\left(a_s(s)\right)^m\, .
\label{eq:FOPT}
\end{eqnarray}

To further utilize the FOPT approach, it is useful to consider the
relation between the coefficients $r_m$ and $d_k$ in more detail.
The goal is to express the coefficients $r_m$ in (\ref{eq:FOPT})
in terms of the calculable coefficients $d_n$ in Eq.\ (\ref{eq:D.Q}),
i.e., to write $r_m=T^{mk}d_k$, where summation over
$k=1,\ldots,m$ is implied.
The matrix $T^{mk}$ is triangular with unity elements on
its diagonal---see Table \ref{Tab:r_n.d_k}.
In the horizontal direction, i.e., along the rows of this Table,
we include all coefficients $d_i$ up to the coefficient $d_5$,
the latter not calculated yet, but due to be estimated within our
approach later in connection with specific
applications---Secs.\ \ref{sec:App.Higgs} and \ref{sec:App.Adler}).

The elements, proportional to $\beta_0^m$, which originate from the
one-loop evolution procedure, have the following general form
\begin{eqnarray}
\label{eq:one-loop.r_n}
  d_n\,\frac{a_s^n}{4^n} \sum_{m=0}^{} \frac{(n-1+2m)!}{(n-1)!~(2m)!}
  (-1)^m \frac{(a_s\,\pi\,\beta_0)^{2 m}}{2m+1}
\end{eqnarray}
and can be obtained for any fixed order $n$ of the expansion by the
procedure described in Appendix \ref{App:A}.
The other $\beta_i$-coefficients---related to higher loops---have
been color-printed below using the same color assignments as in
Table \ref{Tab:r_n.d_k}.

To get acquaintance with the use of this Table, we write out explicitly
the relation between $r_6$ and $d_i$, $i=1,\ldots,6$,
(printing the four-loop-order contribution in green color):\footnote{
The four-loop-order contribution has also been computed
in Ref.\ \cite{KaSt95}.
The two results coincide.}
\begin{eqnarray}
  r_6
& & \!\!\!\! = ~
  \left[
         \bm{\frac{(\pi\,\beta_0)^2}{3}\,0}
        + \RedTn{\bm{\frac{77\,\pi^4}{60}\,\beta_0^3\,\beta_1}}
        - \BluTn{\bm{\frac{7\,\pi^2}{6}\,\beta_1\,\beta_2}}
        - {\DarkGreen\bm{\frac{7\,\pi^2}{6}\,\beta_0\,\beta_3}}
  \right]
         \,d_1
\nonumber\\
& & +
   \left[
          \bm{\frac{(\pi\,\beta_0)^4}{5}\,5}
         -\RedTn{\bm{\frac{4\,\pi^2}{3}\,\beta_1^2}}
         -\BluTn{\bm{\frac{8\,\pi^2}{3}\,\beta_0\,\beta_2}}
   \right]
          \,d_2
\nonumber\\
& & +
   \left[
          \bm{\frac{(\pi\,\beta_0)^2}{3}\,0}
         -\RedTn{\bm{\frac{9\,\pi^2}{2}\,\beta_0\,\beta_1}}
   \right]
          \,d_3
\nonumber\\
& & -
   \left[
         \bm{\frac{(\pi\,\beta_0)^2}{3}\,10}
   \right]
          \,d_4
+
      \left[
            \bm{\frac{(\pi\,\beta_0)^2}{3}\,0}
      \right]
             \,d_5
             + d_6\, .
\label{eq:r_6.d_6}
\end{eqnarray}
The role of the ``kinematical $\pi^2$''-terms becomes more pronounced
for higher orders $m$.
For instance, as regards the term $r_3$, the underlined
$\pi^2$-contributions are comparable in size with the original
$d_3$-contribution, while for $r_4$ these contributions even exceed in
magnitude the value of the coefficient $d_4$ \cite{BCK08,BCK10}, as one
observes from the expressions
\begin{eqnarray}
  r_3
&=&
   18.2 - \underline{24.9} + (-4.22 + \underline{3.02})n_f
  +(-0.086 + \underline{0.091})n_f^2\,,
\\
  r_4
&=&
    135.8 - \underline{292.4} + (-34.4 + \underline{53.2})n_f
   +(1.88 - \underline{2.67})n_f^2
   +(-0.01+\underline{0.032})n_f^3\, .
\end{eqnarray}
Therefore, we may be tempted to take into account the ``kinematical''
$\pi^2$-terms to all orders of the expansion, i.e., to sum, for a
fixed element $d_n$, over the index $m$
\textit{along a single column} in Table \ref{Tab:r_n.d_k}
by taking into account the corresponding power of the coupling
$a_s^m$.

Let us now look at alternative QCD perturbative expansions.

\subsection{Contour-improved perturbation theory}
\label{subsec:CIPT}
\begin{figure}[ht]
 \centerline{\includegraphics[width=0.6\textwidth]{
  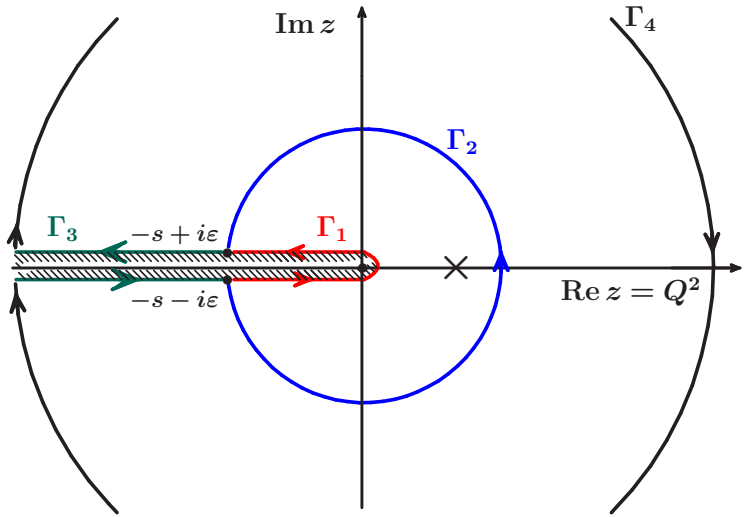}}
 \caption{Integration contour $\Gamma_1$ entering the calculation
  of $R(s)$ in Eq.\ (\protect\ref{eq:R-D-operation}).
  The position of the Landau pole is marked by the cross symbol
  {\Large$\bm{\times}$}. The other integration contours shown
  are explained in the text.\label{fig:G1G2G3}}
\end{figure}

Another way to determine $R$ was suggested in \cite{Piv91eng,LeDiPi92},
where the integration contour in Eq.\ (\ref{eq:R-D-operation}) was
changed from the cut $\Gamma_1$ to the contour $\Gamma_2$,
i.e., to the circle in the complex $Q^2$-plane with radius
$s$ centered at $0$, see Fig.\ \ref{fig:G1G2G3}.
Applying the FOPT expansion, the integration of the terms
$d_n(\sigma/\mu^2)\,a_s^n(\mu^2)$
in Eq.\ (\ref{eq:D.Q}),
along the contour $\Gamma_2$, is completely equivalent to the
integration along the cut.
Using instead the so-called Contour-Improved Perturbation Theory
(CIPT), for large enough values of $s$, one can integrate the terms
$d_n\,a_s^n(\sigma)$, entering its RHS, along the contour
$\Gamma_2$ to get
\begin{equation}
  A_{n}
\equiv
  \frac1{2\pi i}
  \int_{\Gamma_{2}}a_s^{n}(\sigma) \frac{d\sigma}{\sigma}\, .
\label{eq:A_n.CIPT}
\end{equation}
This means that one may absorb all logarithms inside the running
coupling just by adjusting in Eq.\ (\ref{eq:R-D-operation}) the
magnitude of the scale $\mu^2=\sigma$---without performing any
expansion.
CIPT is currently considered to be the most preferable technique
to account for the running of perturbative quantities in a number of
processes, including the $\tau$-decay
\cite{GKP01,DDGHMZ08},
the $R_{e^+e^-}$-ratio \cite{BCK08,BCK10},
and the width of the Higgs-boson decay H$\to b\bar{b}$ \cite{KK09}.
The integration of the running couplings along the contour $\Gamma_2$
is performed numerically \cite{DDGHMZ08}, but for the one-loop running
the result for $R^{\text{CIPT}}$ can be obtained also analytically
and is found to coincide with expression (\ref{eq:one-loop.r_n}),
as expected (see Appendix \ref{App:A}).

\subsection{Fractional analytic perturbation theory}
\label{subsec:FAPT}
Inspection of Table \ref{Tab:r_n.d_k} suggests to consider the sum of
the elements of each of its columns as defining a new coupling.
For instance, the first column, associated with $d_1$, gives rise to
the coupling ${\mathfrak A}_{1}$, while the second column pertains to
the coupling ${\mathfrak A}_{2}$, and so on.
In this way, one becomes able to introduce a \textit{non-power}-series
expansion in Minkowski space.
This is, actually, just another way to define the Analytic Perturbation
Theory \cite{SS97,MS96,SS06}, and its extension---Fractional Analytic
Perturbation Theory \cite{BMS05,BMS06}
(see also \cite{BKS05,KS01} and \cite{AB08gfapt,Ste09} for reviews).
The basis of (F)APT is provided by the following linear operations
which define analytic images of the normalized coupling
$a=a_s\,b_0/4$ and its powers
in the Euclidean and the Minkowski space,
respectively:\footnote{To streamline our notation, we use for all quantities
in the  Euclidean space a calligraphic symbol, whereas their
Minkowski-space counterparts are denoted by Gothic symbols.
Moreover, note that Latin indices represent integer numbers $n$,
whereas Greek indices $\nu$ mark real numbers.
The subscripts E and M serve to specify the space we are
working in: Euclidean and Minkowski, respectively.}
\begin{equation}
  {\mathcal A}_{\nu}(Q^2)
=
  \textbf{A}_\text{E}\left[a^{\nu}_{}\right]
\equiv
  \int_0^{\infty}\!
  \frac{\rho_\nu(\sigma)}{\sigma+Q^2}\,
  d\sigma\, , ~~~~~
  {\mathfrak A}_{\nu}(s)
=
  \textbf{A}_\text{M}\left[a^{\nu}_{}\right]
\equiv
  \int_s^{\infty}\!
  \frac{\rho_\nu(\sigma)}{\sigma}\,
  d\sigma~,
\label{eq:A.U.rho.Nf}
\end{equation}
where
$
 \Ds \rho_\nu(\sigma)=\frac1{\pi}\textbf{Im}\,
 \left[a^{\nu}(-\sigma)\right]
$
is the spectral density.
The set of the couplings
$\{{\mathcal A}_{\nu},\ {\mathfrak A}_{\nu} \}$
satisfy the dispersion relations Eq.\ (\ref{eq:R-D-operation}) and
fulfill the constraint $\hat{R} \hat{D} = \1 $.
Applying the operation $\textbf{A}_\text{M}$ on the RHS of the last
expression in Eq.\ (\ref{eq:D.Q}), one obtains $R$ in APT:
\begin{eqnarray}
  \textbf{A}_\text{M}[D]
=
  R^\text{APT}
=
  d_0+\sum_{m=1}\hat{d}_m {\mathfrak A}_{m}\, .
\label{eq:MFAPT}
\end{eqnarray}
The reader should note here that the expansion coefficients $d_n$
differ from those in Eq.\ (\ref{eq:D.Q}) and are defined as follows
\begin{eqnarray}
\label{eq:hat-d_n}
  \hat{d}_n
&=&
  \left(\frac{4}{b_0}\right)^n d_n\, .
\end{eqnarray}
The prefactors $(4/b_0)^n$ above---and analogously in
Eq.\ (\ref{eq:minkcouplexp}) below--- serve to connect this analysis
to our standard definitions of the analytic couplings
$\mathcal A_\nu$ and $\mathfrak A_\nu$
in Refs.\ \cite{BMS05,BMS06}.
Hence, according to Eqs.\ (\ref{eq:A.U.rho.Nf})--(\ref{eq:MFAPT}),
one may associate the evaluation of $R$ within (F)APT with the
integration contour $\Gamma_3$ in Fig.\ \ref{fig:G1G2G3}.

Consider now how the (F)APT result $R^\text{APT}$ in Eq.\ (\ref{eq:MFAPT})
correlates with the elements of Table \ref{Tab:r_n.d_k}.
One appreciates that every term $\hat{d}_m\,{\mathfrak A}_{m}$
of the series appears as an \textit{infinite} sum of the elements
along the corresponding matrix column in this table.
The series is convergent and includes all so-called
``kinematical terms'', as we discussed at length in \cite{BMS06}.
One can verify this by considering the expansion of the first few
elements $\hat{d}_m\,{\mathfrak A}_{m}$
in Eq.\ (\ref{eq:minkcouplexp})
and then compare the results with the content of the corresponding
column in Table \ref{Tab:r_n.d_k}.
A further advantage of Eq.\ (\ref{eq:minkcouplexp}) is that one can use
it to derive expression (\ref{eq:one-loop.r_n}) in a more direct way,
namely, as an expansion of the one-loop Minkowski analytic couplings
${\mathfrak A}_{n}$
in powers of the variable $a_s$---terms printed in black color
in Eq.\ (\ref{eq:minkcouplexp}).
The analogous terms for higher loops can be found in Appendix C
of Ref.\ \cite{BMS06}.
Thus, we obtain
(with the same coloring assignments as in Sec.\ \ref{subsec:FOPT})
\begin{subequations}
\label{eq:minkcouplexp}
\begin{eqnarray}
  \left(\frac{4}{b_0}\right)^{1}{\mathfrak A}^{(3)}_{1}
= \!\!&\!\!&\!\!\!\!
  a_s \left[1-\frac{(a_s \pi \beta_0)^2}{3}
            +\frac{(a_s \pi \beta_0)^4}{5}
            +\ldots
      \right]
\nonumber \\
  \!\!&\!\RedTn{\bm{-}}\!&\!\!
         \RedTn{\bm{\frac{\pi^2}{3}\,a_s^4\,
               \Big(\beta_0\beta_1\frac{5}{2}
                  + \beta_1^2\frac{3}{2} a_s
                  - \frac{\pi^2}{5}\beta_0^3\beta_1\frac{77}{4}a_s^2
                  + \ldots
               \Big)}
               }
\nonumber \\
  \!\!&\!\BluTn{\bm{-}}\!&\!\!
         \BluTn{\bm{\frac{\pi^2}{3}\,a_s^5\,
               \left(\beta_0\beta_2~3
                    +\beta_1\beta_2~\frac{7}2 a_s
                    +\ldots
               \right)
                    }
                }
\nonumber\\
\!\!&\!{\DarkGreen\bm{-}}\!&\!\!
         {\DarkGreen\bm{\frac{\pi^2}{6}\,a_s^6\,\beta_0\beta_3\,7}}
         +\ldots
\\
  \left(\frac{4}{b_0}\right)^2{\mathfrak A}^{(2)}_{2}
= \!\!&\!\!&\!\!\!\!
    a^2_s \left[1-(a_s \pi \beta_0)^2 + (a_s \pi \beta_0)^4~\ldots
          \right]
\nonumber \\
  \!\!&\!\RedTn{\bm{-}}\!&\!\!
         \RedTn{\bm{\frac{\pi^2}{3}\,a^5_s\,
                    \left(\beta_0 \beta_1 7+ \beta_1^2 4 a_s+\ldots
                    \right)
                    }
                }
\nonumber \\
\!\!&\!\BluTn{\bm{-}}\!&\!\!
        \BluTn{\bm{\frac{\pi^2}{3}\,a_s^6\,\beta_0\,\beta_2\,8}
              }    + \ldots \\
  \left(\frac{4}{b_0}\right)^3 {\mathfrak A}^{(2)}_{3}
= \!\!&\!\!&\!\!\!\!
  a^3_s \left[1-\frac{(a_s \pi \beta_0)^2}{3} 6~+ \ldots
        \right]
\nonumber \\
\!\!&\!\RedTn{\bm{-}}\!&\!\!
       \RedTn{\bm{\frac{\pi^2}{3}\,a_s^6\,\beta_0\beta_1\frac{27}{2}}}
                  + \ldots
\end{eqnarray}
\begin{eqnarray}
  \left(\frac{4}{b_0}\right)^4 {\mathfrak A}^{(2)}_{4}
=
  a^4_s \left[1-\frac{(a^2_s\pi \beta_0)^2}{3}10+\ldots
        \right]
              + \ldots
\end{eqnarray}
\end{subequations}
The main conclusion from the above exposition is that, depending on
the particular scheme of the perturbative expansion used, the
elements of Table \ref{Tab:r_n.d_k} may be summed in different ways
(see Table \ref{tab:PT-relations}).
Specifically,
\begin{enumerate}
  \item[(i)] FOPT gives the sum of a finite number of terms
   along some row to create $r_m$,
   and then---following Eq.\ (\ref{eq:FOPT})---it sums
   the results up to $n$ to yield $R_{n}$,
   i.e.,
   $R_n=d_0 +\sum\limits^{n}_{m,k=1}a_s^m\,T^{mk}\,d_k$.
  \item[(ii)] (F)APT takes into account each infinite column
   \textit{as a whole}
   in the form of the expansion
   $\hat{d}_m\,{\mathfrak A}_{m}$,
   thus including this way all ``kinematical terms'' by construction,
   and then sums a number of
   ${\mathfrak A}_{m}$ terms into $R^\text{APT}_n$
   in the form given by expression (\ref{eq:MFAPT}).
\end{enumerate}
It is evident that for any fixed order $n$, the results for $R_n$ and
$R^\text{APT}_{n}$ cannot coincide.
Nevertheless, it was shown in \cite{BMS06} that calculating the decay
width of a Higgs boson into a $b\bar{b}$ pair using CIPT, leads to
the same result one would obtain for $R$ using FAPT with one-loop
running of the coupling.
This coincidence turns out to be not accidental but to hold at any
loop order of the perturbative expansion by virtue of the dispersion
relation given in Eq.\ (\ref{eq:A.U.rho.Nf}), as we will prove next.
\begin{table}[h]
\caption{QCD observable $R$ calculated with different perturbative
expansions: (F)APT---(Fractional) Analytic Perturbation Theory;
CIPT---Contour-Improved Perturbation Theory;
FOPT---Fixed-Order Perturbation Theory.
The associated contours $\Gamma_1$, $\Gamma_2$, $\Gamma_3$ are
displayed in Fig.\ \protect\ref{fig:G1G2G3}.
\label{tab:PT-relations}
\vspace*{+3mm}}
\begin{tabular}{|ccc|}\hline
~Perturbative scheme~&~Contours  ~& ~$R$ expansion~\\ \hline
~FOPT~               &~$\Gamma_1$~& ~$\sum_{n} r_{n}a_s^{n}$                \\
~CIPT~               &~$\Gamma_2$~& ~$\sum_{n} d_{n}A_{n}$                  \\
~(F)APT\,\footnote{%
 The coefficients $d_{n}$ and $\hat{d}_{n}$ in CIPT and (F)APT differ
 by trivial factors, see Eq.\ (\ref{eq:hat-d_n}),
 due to the different normalization of the couplings $A_{n}$
 and $\mathfrak{A}_{n}$.}
                     &~$\Gamma_3$~& ~$\sum_{n} \hat{d}_{n}\mathfrak{A}_{n}$ \\
\hline\end{tabular}
\end{table}

\subsection{Fixed flavor number vs. global FAPT}
\label{sec:Fix-Nf.Global.APT}
We commence our analysis within FAPT by recalling the salient features
of the analytic approach to QCD perturbation theory, expanding our
remarks given in Section \ref{subsec:FAPT}.
In order to have a direct connection to our previous papers on the
subject \cite{BMS05,BKS05,BMS06}, and to simplify the main formulae
in those sections where we consider fixed-order (F)APT with a constant
value of active flavors $N_f$, we use here the normalized coupling of
perturbative QCD (pQCD) \cite{MS04}
$a=\beta_f\,\alpha_\text{s}$
with the useful abbreviation
$
 \beta_f\equiv b_0(N_f)/(4\,\pi)
=
 (11-2\,N_f/3)/(4\,\pi)
$,
where $b_0(N_f)$ denotes the first coefficient of the QCD
$\beta$ function.
Analytic images of the normalized coupling and its powers are
constructed by means of the linear operations
$\textbf{A}_\text{E}$ and $\textbf{A}_\text{M}$ according to
(\ref{eq:A.U.rho.Nf}) using the spectral density
$
 \Ds\rho_\nu(\sigma)
=
 \frac1{\pi}\textbf{Im}\,\left[a^{\nu}(-\sigma)\right]
$.

To be in line with the above definitions, we also introduce analogous
expressions for the fixed-$N_f$ quantities with standard normalization,
i.e.,
\begin{eqnarray}
 \bar{\mathcal A}_{\nu}(Q^2;N_f)
=
  \frac{{\mathcal A}_{\nu}(Q^2)}{\beta_{f}^{\nu}}\, , \quad
  \bar{\mathfrak A}_{\nu}(s;N_f)
=
  \frac{{\mathfrak A}_{\nu}(s)}{\beta_{f}^{\nu}}\, ,
\label{eq:Bar.Couplings}
\end{eqnarray}
which correspond to the analytic couplings
${\mathcal A}_{\nu}$ and ${\mathfrak A}_{\nu}$
in the Shirkov--Solovtsov terminology \cite{SS06}.
These couplings have dispersive representations of the type
(\ref{eq:A.U.rho.Nf}) with spectral densities
$
 \bar{\rho}_\nu(\sigma;N_f)
=
 {\rho}_\nu(\sigma)/{\beta_f^\nu}
$.
For the sake of simplicity we will omit to display $N_f$
(and other evident arguments) explicitly.
In the present analysis we express all variables in terms of
$L=\ln(Q^2/\Lambda^2)$ (Euclidean space)
or $L=\ln(s/\Lambda^2)$ (Minkowski space),
using the notation defined above, but with an argument placed in
square brackets:
$a^\nu[L]$, ${\mathcal A}_\nu[L]$, and ${\mathfrak A}_\nu[L]$.
Then, in the one-loop approximation
(labeled by the superscript (1)),
we have
\begin{subequations}
\label{eq:couplings.one-loop}
\begin{eqnarray}
&& a^{(1)}[L]
=
  \frac{1}{L}\,,  \quad {\mathcal A}^{(1)}_{1}[L]
=
  \frac1{L}- \frac1{e^L -1}\, , \quad
  {\mathfrak A}^{(1)}_{1}[L]
=
  \frac{1}{\pi}\,
               \arccos\left(\frac{L}{\sqrt{L^2+\pi^2}}\right)\, ,
\label{eq:a.one-loop}\\
&&
  \rho^{(1)}_1[L]
=
  \frac{1}{L^2+\pi^2}\,, \quad
  \bar{\rho}^{(1)}_1[L;N_f]
=
  \frac{1}{\beta_f\left(L^2+\pi^2\right)}\, .
\label{eq:rho.one-loop}
\end{eqnarray}
\end{subequations}
On the other hand, when we will discuss the global version of (F)APT,
where $Q^2$ (or $s$) varies in the whole (``global'') domain
$[0,\infty)$, and $N_f$ effectively becomes dependent on $Q^2$
(or $s$), we will use in Eq.\ (\ref{eq:A.U.rho.Nf}) the global version
of the spectral density that takes into account threshold effects and
is, therefore, $N_f$-dependent.

In order to make the effect of crossing a heavy-quark threshold
more plausible, consider a single threshold at $\sigma=m_4^2$
(corresponding to the transition $N_f=3 \Rightarrow N_f=4$)
and write the spectral density in a form which expresses it in terms
of the fixed-flavor spectral densities with 3 and 4 flavors,
$\bar{\rho}_n[L;3]$ and $\bar{\rho}_n[L+\lambda_4;4]$, respectively.
The result is
\begin{eqnarray}
\label{eq:global_PT_Rho_4}
  \rho_\nu^\text{\tiny glob}(\sigma)
=
  \rho_\nu^\text{\tiny glob}[L_\sigma]
=
  \theta\left(L_\sigma<L_{4}\right)\,
  \bar{\rho}_\nu\left[L_\sigma;3\right]
  + \theta\left(L_{4}\leq L_\sigma\right)\,
  \bar{\rho}_\nu\left[L_\sigma+\lambda_4;4\right]~~~
\end{eqnarray}
with $L_{\sigma}\equiv\ln\left(\sigma/\Lambda_3^2\right)$,
$L_{4}\equiv\ln\left(m_4^2/\Lambda_3^2\right)$,
and
$\lambda_4\equiv\ln\left(\Lambda_3^2/\Lambda_4^2\right)$.
Note, however, that in standard QCD perturbation theory the coupling at
the threshold becomes discontinuous.
To enable a smooth matching at this point, one has to readjust the
values of $\Lambda_\text{QCD}$ on each side of the threshold in
correspondence with their associated flavor numbers
($\Lambda_{f}=\Lambda_\text{QCD}^{N_f=f}$).
\begin{figure}[b!]
 \centerline{\includegraphics[width=0.45\textwidth]{
  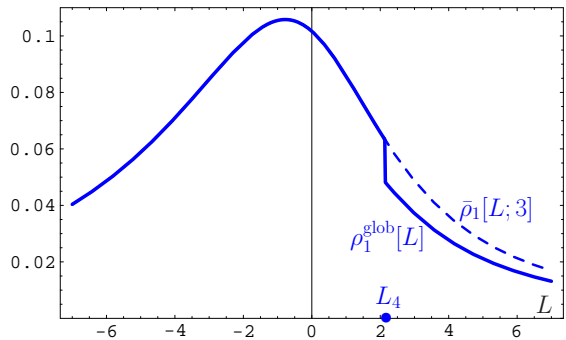}\vspace*{-1mm}}
  \caption{Spectral densities $\rho_1^\text{\tiny glob}[L]$
   (solid line)
   and $\bar{\rho}_1[L;3]$ (dashed line).
   The discontinuity at $L=L_4$ is artificially enhanced by a factor
   of 1.5 to make it more visible.
   \label{fig:SpDen.34}}
\end{figure}
In Fig.\ \ref{fig:SpDen.34} we show the global spectral density
$\rho_1^\text{\tiny glob}[L]$
in comparison with the spectral density $\bar{\rho}_1[L;3]$ which
corresponds to a fixed flavor number $N_f=3$.
We see that, because of the different values of $\Lambda_3$ and
$\Lambda_4$, and also of $\beta_3$ and $\beta_4$, these quantities
differ from each other when $L$ crosses $L_4$.

\begin{figure}[h!]
 \begin{minipage}{\textwidth}
  \centerline{\includegraphics[width=0.43\textwidth]{
   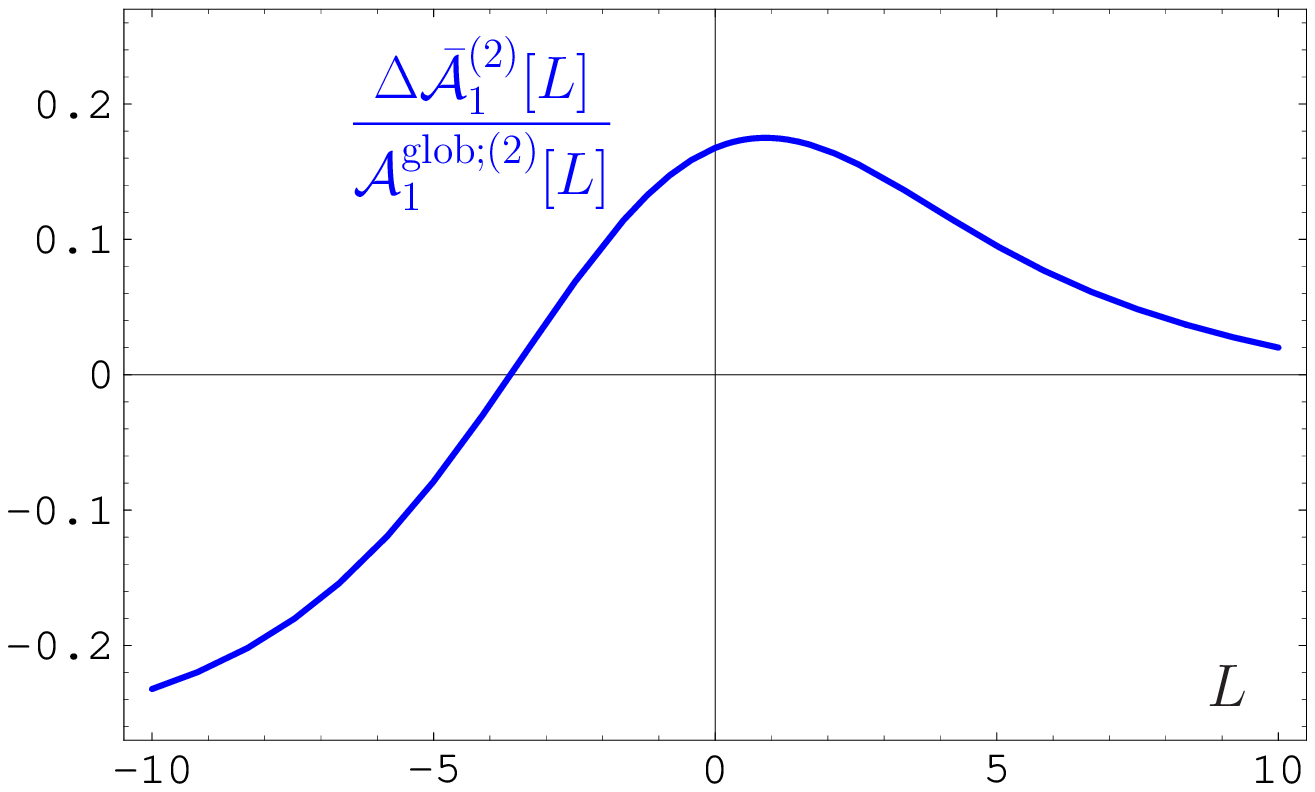}~~~\includegraphics[width=0.43\textwidth]{
   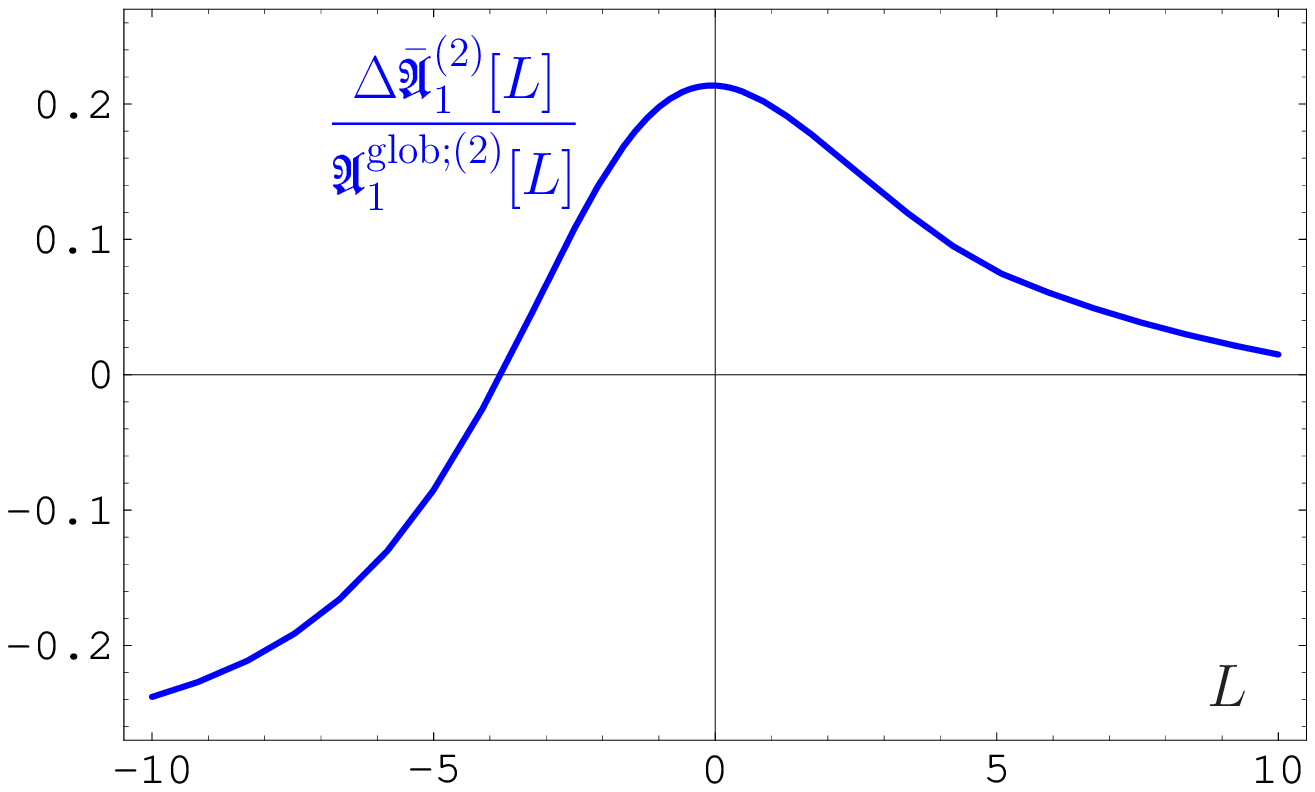}\vspace*{-1mm}}
   \caption{\small Left: Deviation of the global coupling relative
   to the fixed-$N_f$ coupling in FAPT:
   $\Delta\overline{\mathcal A}_{1}[L]/\mathcal A_{1}^\text{\tiny glob}[L]$.
   Right: The same for
   $\Delta\overline{\mathfrak A}_{1}[L]/\mathfrak A_{1}^\text{\tiny glob}[L]$.
\label{fig:Delta.Global.A_1}}
\end{minipage}
\end{figure}

Substituting the obtained spectral density
$\rho_\nu^\text{\tiny glob}(\sigma)$
[cf.\ Eq.\ (\ref{eq:global_PT_Rho_4})]
into Eq.\ (\ref{eq:A.U.rho.Nf}),
we get continuous expressions for the analytic couplings in both
domains of the complex $Q^2$ space.
In the Minkowski region, the global analytic coupling reads
\begin{eqnarray}
  {\mathfrak A}_{\nu}^{\text{\tiny glob}}[L]
= \theta\left(L\!<\!L_4\right)
  \Bigl\{\bar{{\mathfrak A}}_{\nu}^{}[L;3]
        -\bar{{\mathfrak A}}_{\nu}^{}[L_4;3]
        +\bar{{\mathfrak A}}_{\nu}^{}[L_4+\lambda_4;4]
  \Bigr\}
  + \theta\left(L_4\!\leq\!L\right)\,
        \bar{{\mathfrak A}}_{\nu}^{}[L+\lambda_4;4]\, ,~~~
\label{eq:An.U_nu.Glo.Expl}
\end{eqnarray}
whereas its Euclidean counterpart assumes the form (referring for more
details to \cite{BM08,AB08gfapt})
\begin{eqnarray}
  {\cal A}_{\nu}^{\text{\tiny glob}}[L]
&=& \bar{{\cal A}}_{\nu}^{}[L+\lambda_4;4]
  + \int\limits_{-\infty}^{L_4}\!
  \frac{\bar{\rho}_{\nu}^{}[L_\sigma;3]
       -\bar{\rho}_{\nu}^{}[L_\sigma+\lambda_{4};4]}
       {1+e^{L-L_\sigma}}\,
       dL_\sigma\, .
\label{eq:Delta_f.A_nu}
\end{eqnarray}
To demonstrate the magnitude of the threshold corrections, we show in
Fig.\ \ref{fig:Delta.Global.A_1} the values of the normalized
deviations
$\Delta\overline{\mathcal A}_{\nu}[L]=
 {\mathcal A}_{\nu}^{\text{\tiny glob}}[L]
 -\overline{\mathcal A}_{\nu}[L\!+\!\lambda_4;4]
$
and
$\Delta\overline{\mathfrak A}_{\nu}[L]=
 {\mathfrak A}_{\nu}^{\text{\tiny glob}}[L]
 -\overline{\mathfrak A}_{\nu}[L\!+\!\lambda_4;4]
$
in the Euclidean and the Minkowski domain,
respectively.
We see that in both domains these deviations vary from $-20\%$ for
large values of $-L\approx10$, going through zero in the vicinity of
$L\approx-5$,
and then increase up to the value $+20\%$ for $L\approx0$,
tending, finally, to 0 as $L\to\infty$.
This means that these deviations reach the level of 10\% in the region
of several tens of GeV$^2$.

\subsection{Relation between CIPT and (F)APT}
\label{subsec:CIPT.FAPT}
To establish the equivalence between FAPT and CIPT, we consider a more
general expansion than Eq.\ (\ref{eq:D.Q}) which contains the coupling
with a non-integer power that can be related to an anomalous dimension.
Such a quantity reads
\begin{eqnarray}
  D_{\nu}
=
  d_0 a_{s}^{\nu} + \sum_{n=1} d_n a_{s}^{\nu+n}\,,
\end{eqnarray}
where $\nu$ is not an integer number.

Symbolically, we have the following equivalence
\begin{equation}
 \text{FAPT} \left(\int_{\Gamma_{3}}\right)
=
  \text{CIPT} \left(\int_{\Gamma_{2}}\right)\,,
\label{eq:equiv}
\end{equation}
referring for the integration contours to Fig.\ \ref{fig:G1G2G3}.
To establish the equivalence between FAPT and CIPT in the above
relation, we employ the chain of the equalities
\begin{equation}
  \text{FAPT}
\equiv
  \int_s^{\infty}\!
  \frac{\rho_\nu(\sigma)}
  {\sigma}\,
  d\sigma
\equiv
  \int_s^{\infty}\!
  \frac{\textbf{Im}\,\left[a^{\nu}(-\sigma)\right]}{\pi}
  \frac{d \sigma}{\sigma}\
=
  - \frac1{2\pi i} \int_{\Gamma_{3}}a^{\nu}(\sigma) \frac{d\sigma}
  {\sigma}\, ,
\label{eq:theorem-1}
\end{equation}
where in the last step the integral has to be evaluated along the
contour $\Gamma_{3}$.
On the other hand, the CIPT part of Eq.\ (\ref{eq:equiv}) can be
identically rewritten as
\begin{equation}
  \text{CIPT}
\equiv
  \frac1{2\pi i}
  \int_{\Gamma_{2}}a^{\nu}(\sigma) \frac{d\sigma}{\sigma}\, .
\label{eq:theorem}
\end{equation}
Therefore, we have to prove that
\begin{equation}
  \int_{\Gamma_{3}}\left(a(\sigma)\right)^{\nu}
  \frac{d\sigma}{\sigma}
+  \int_{\Gamma_{2}}\left(a(\sigma)\right)^{\nu}
  \frac{d\sigma}{\sigma}=0\,.
\label{eq:tho-contours}
\end{equation}
To this end, we close the contour $\Gamma_{3}$ along the large
circle $\Gamma_{4}$ with radius $R$ that tends to $\infty$,
and take into account the closed composed contour
$\Gamma_{234}=\Gamma_2\cup\Gamma_3\cup\Gamma_4$
(see Fig.\ \ref{fig:G1G2G3}).
The integral $I(\Gamma_{234})$,
\begin{equation}
  I(\Gamma_{234})
=
  \frac1{2\pi i} \oint_{\Gamma_{234}}a^{\nu}(\sigma)
  \frac{d\sigma}{\sigma}\, ,
\label{eq:Gamma.234}
\end{equation}
along the closed contour $\Gamma_{234}$ is equal to the sum of the
residues inside the enclosed region.
Provided the radius $s$ of the contour $\Gamma_2$ is large enough,
no poles from $a(\sigma)$ owing to perturbation theory are inside
the contour $\Gamma_{234}$.
As a result, $I(\Gamma_{234})=0$.
Moreover,
$(a(\sigma))^\nu$ ($\nu>0$)
on the contour $\Gamma_4$ decreases with growing $R$, and therefore
this contribution vanishes as $R \to \infty$.
Consequently, we finally obtain
\begin{equation}
 I(\Gamma_{234}) \stackrel{R\to\infty}{\longrightarrow}
 I(\Gamma_2\cup\Gamma_3)
= \frac1{2\pi i} \int_{\Gamma_2\cup\Gamma_3}a^{\nu}(\sigma)
    \frac{d\sigma}{\sigma}
= 0 \, .
\label{eq:residues}
\end{equation}

The above equivalence not withstanding, there is a crucial advantage
of the FAPT approach with respect to CIPT.
In fact, as long as one is only interested in a numerical estimate,
CIPT provides acceptable results for several typical processes.
However, if one pretends to employ \emph{analytic} expressions, and
thereby to control each step of the calculation, CIPT is not sufficient.
In that case, one needs another perturbative scheme that is able to
yield explicit expressions for the couplings along each column
in Table \ref{Tab:r_n.d_k}.
Such a scheme is naturally provided by (F)APT.
Moreover, we shall show below that this scheme can even admit the
resummation over columns \cite{BM08}.
This is an important feature, given that the conventional perturbative
series in Eq.\ (\ref{eq:D.Q}) cannot amount to a unique, i.e.,
resummation-method-independent, result owing to the asymptotic
nature of the power series.

\section{Resummation in (F)APT in the one-loop order}
\label{sec:APT.Resummation}
In this section, we consider different sorts of perturbation-series
expansions of typical physical quantities, like the Adler function,
$D[L]$.
Our goal is to perform the summation of such expansions under the
imposition of a couple of basic constraints.
As we will show in a moment, these constraints are
(i) a recurrence relation for higher-order couplings and
(ii) the nonpower character of the series.
These considerations will be based on an appropriate generating
function for the expansion coefficients.

In what follows we discuss in detail the one-loop running case.
However, also the technically more complicated two-loop running case
is worked out and the corresponding expressions are provided
in Appendix \ref{app:Two-loop}.

\subsection{Generating function for the series expansion}
\label{subsec:P(t)}
At the one-loop level \cite{BMS06}, we have
\begin{eqnarray}
\label{eq:Ini.Series}
  \left\{
\begin{array}{l}
  D[L]\\
{D}^\text{\tiny APT}[L]\\
 {R}^\text{\tiny APT}[L]
\end{array}
  \right\}
=
  d_0+\hat{d}_1\sum_{n=1}^{\infty} \tilde{d}_n\,
  \left\{
\begin{array}{l}
  a^{n}[L]\\
  {\mathcal A}_{n}[L]\\
  {\mathfrak A}_{n}[L]
  \end{array}
  \right\}
\end{eqnarray}
with
$\tilde{d}_n\equiv \hat{d}_n/\hat{d}_1$
[cf.\ Eq.\ (\ref{eq:hat-d_n})],
where
$L=\ln\left(Q^2/\Lambda^2\right)$
applies to the Euclidean quantities
($D[L]$, ${D}^\text{\tiny APT}[L]$, $a^n[L]$, and ${\mathcal A}_n[L]$),
and
$L=\ln\left(s/\Lambda^2\right)$
pertains to their Minkowski versions
${R}^\text{\tiny APT}[L]$ and ${\mathfrak A}_n[L]$.

The resummation of this series is on the focus of the present work.
For this purpose, it is useful to introduce a generating function
$P(t)$ for the series expansion and write
\begin{equation}
  \tilde{d}_n
=
  \int_{0}^\infty\!\!P(t)\,t^{n-1}dt
  ~~~\text{with}~~~
  \int_{0}^\infty\!\!P(t)\,d t = 1\, .
\label{eq:generator}
\end{equation}
We will show in the next step how to use this generating function in
order to resum nonpower series expansions
(both in the Euclidean and the Minkowski region).
But first recall that the standard coupling $a^n$ of perturbative QCD,
as well as the couplings
$ {\mathcal A}_{n}, {\mathfrak A}_{n}$---together with the spectral
density $\rho_n$---satisfy a one-loop renormalization-group equation
that can be recast in the form of a recurrence relation \cite{Shi98}:
\begin{equation}
  \left\{
\begin{array}{l}
  a^{n+1}[L]\\
  {\mathcal F}_{n+1}[L]\\
\end{array}
  \right\}
=
  \frac{1}{\Gamma(n+1)}\left(-\frac{d}{d L}\right)^{n}
  \left\{
  \begin{array}{l}
  a^{1}[L]\\
  {\mathcal F}_{1}[L]\\
\end{array}
  \right\}\, .
\label{eq:recurrence}
\end{equation}
Here ${\mathcal F}[L]$ denotes one of the analytic quantities
${\mathcal A}[L],~{\mathfrak A}[L],~\rho[L]$.
Substituting Eqs.\ (\ref{eq:generator}) and (\ref{eq:recurrence})
into the perturbative-series expansion, i.e., into Eq.\
(\ref{eq:Ini.Series}), one obtains \cite{MS04} in analogy to the
previous equation,
\begin{eqnarray}
  \left\{
  \begin{array}{l}
  D[L]\\
  {\mathcal S}[L]\\
\end{array}
  \right\}
=
  d_0
     + \hat{d}_1\sum_{n=0}^{\infty}
         \frac{\left\langle\!\left\langle{(-t)^{n}}\right\rangle\!
               \right\rangle_{P}}
               {n!}\,
         \frac{d^n}{dL^n}
         \left\{
\begin{array}{l}
         a[L]\\
         {\mathcal F}_{1}[L]\\
\end{array}
         \right\}
=
  d_0 + \hat{d}_1
  \left\{
\begin{array}{l}
             \left\langle\!\left\langle{a[L-t]}\right\rangle\!
             \right\rangle_{P}\\
             \left\langle\!\left\langle{{\mathcal F}_{1}[L-t]}
             \right\rangle\!\right\rangle_{P}\\
\end{array}
  \right\}\, ,~~~
\label{eq:APT.Sum}
\end{eqnarray}
where
$\mathcal S=D^\text{\tiny APT}$ or $R^\text{\tiny APT}$---depending
on the choice of the analytic couplings
$\mathcal F=\mathcal A$ or $\mathcal F=\mathfrak A$,
respectively.
In the above equation, and in the considerations to follow,
we use the abbreviated notation
\begin{eqnarray}
  \left\langle\!\left\langle{ {\mathcal F}[L-t]}
                     \right\rangle\!
  \right\rangle_{P}
\equiv
  \int^{\infty}_{0}{\mathcal F}[L-t] P(t)~dt\, .
\label{eq:convolution}
\end{eqnarray}

As long as we have not proved that summation and integration can be
interchanged, this representation has only a formal meaning.
Note, however, that the integration over the Taylor-series expansion
of the term $a[L-t]$ in the integrand reproduces the initial
series for any partial sum.
The integrand in the standard case of the QCD running coupling
(first line in Eq.\ (\ref{eq:APT.Sum}))
faces a pole singularity (termed the infrared renormalon singularity)
and is, therefore, ill-defined.
In contrast, the integral in the second entry in
Eq.\ (\ref{eq:APT.Sum}) has a rigorous meaning by virtue of the
finiteness of $\mathcal F_{1}[L]$, being one of the quantities
$
 \{\mathcal A_{1}[L],\
 \mathfrak A_{1}[L]  \leq 1 ,\
 \rho_{1}[L] \leq  1/\pi^2 \}
$.
Therefore, the expression on the RHS of Eq.\ (\ref{eq:APT.Sum}),
together with (\ref{eq:convolution}), can be called the sum of the
corresponding series in the sense of Euler.
Since any coefficient $\tilde{d}_n$ is the moment of $P(t)$
[cf.\ Eq.\ (\ref{eq:generator})],
this function should fall off faster than any power---e.g., like an
exponential or faster.\footnote{
Let us mention in this context that an expression similar to
(\ref{eq:APT.Sum}) was also obtained in \cite{CvVa06}.
The authors of this paper have used the ``large $\beta_0$'' approximation
to create a specific model
for the generating function
in conjunction with Neubert's approach \cite{Neu95prd}.
}
Therefore, all APT expressions on the RHS of Eq.\ (\ref{eq:APT.Sum})
(second line) exist and are proportional to
${\mathcal F}[L-\bar{t}(L)]$,
where, for each $L$, $\bar{t}(L)$ is the average value of $t$
associated with this quantity.

Provided the generating function is known, one can compute the integral
(\ref{eq:convolution}) explicitly and obtain this way all-order
estimates for the expanded quantity for any series expansion,
provided the coupling parameters fulfill the following conditions:\\
(i) satisfy the one-loop renormalization-group equation
[cf.\ Eq.\ (\ref{eq:recurrence})], \\
(ii) are real, and \\
(iii) have only integrable singularities. \\
If the first coefficient $d_0$ of the expansion (\ref{eq:Ini.Series})
is not accompanied by unity ($a^0$), but by some other fractional power
$\nu$ of the coupling ($a^\nu$), then, as it has been shown in
\cite{BM08,AB08gfapt}, the resummation method (\ref{eq:APT.Sum}) has
to be modified to read
\begin{eqnarray}
  \mathcal S_\nu[L]
=
  d_0\,\mathcal F_\nu[L]
  + \hat{d}_1 \left\langle\!\left\langle{{\mathcal F}_{1+\nu}[L-t]}
                           \right\rangle\!
              \right\rangle_{P_\nu}\, ,
\label{eq:FAPT.Sum}
\end{eqnarray}
where now the generating function $P_\nu$ depends also on $\nu$.
This quantity can be deduced from $P$ as follows:
\begin{eqnarray}
  P_{\nu}(t)
=
  \int_0^{1}\!P\left(\frac{t}{1-x}\right)
  \Phi_{\nu}(x) \frac{dx}{1-x}\, .
\label{eq:P.nu}
\end{eqnarray}
Here
$\Phi_{\nu}(x) = \nu x^{\nu-1}$, so that
$\lim_{\nu\to 0}\Phi_{\nu}\rightarrow \delta(x)$, and therefore
$\lim_{\nu\to 0}P_{\nu}(t)= P(t)$.
The last step completes the generalization of the original
APT resummation procedure of \cite{MS04} to the case of FAPT.

\subsection{Modeling the expansion coefficients and their generating
            function}
\label{sec:model.P(t)}
For most relevant QCD processes, only the first few coefficients
$d_{n}$ are known, while the computation of higher-order coefficients
is technically a highly complicated task.
This is despite the impressing development of sophisticated algorithms
during the last few years \cite{BCK02,BCK05,BCK08,BCK10,BeJam08}.
In view of this, it is extremely useful to have alternative methods
for calculating the higher-order coefficients, let alone to resum the
whole series---even if the result represents a sort of
approximation---provided the quality of the applied method is high
and the inherent uncertainties entailed can be kept under control.
On the other hand, the \textit{asymptotic} form of the coefficients
$\hat{d}_n$ can be predicted from
\begin{eqnarray}
\label{eq:d_n.BLM}
  \hat{d}_n
\sim
  \Gamma(n+1)~n^\gamma c^{n}\left[1+O(1/n)\right]
  \to P(t) \sim
  (t/c)^{\gamma +1}e^{-t/c}
\end{eqnarray}
a form that is inspired by Lipatov's asymptotic expression
in Ref.\ \cite{Lip76} (see also \cite{KS80}, and \cite{Fisch97} for
a review), and where $\gamma<1$ and $c$ are numerical coefficients.
One anticipates that the large-order behavior of the expansion
coefficients translates into the asymptotic form of the generating
function $P(t)$.

To proceed, we have to construct first a model for the expansion
coefficients $\tilde{d}_n=\hat{d}_n/\hat{d}_1$, having recourse to the
information about the first few fixed-order coefficients.
The second step is to interpolate between the obtained result and the
expression for the tail obtained for $\hat{d}(n)$ from
(\ref{eq:d_n.BLM}) at asymptotically large orders $n$.
For a fixed-sign series we adopt the model
(note the normalization $\tilde{d}^\text{model}_1=1$)
\begin{eqnarray}
  \tilde{d}^\text{model}_n
&=&
  \frac{A_1 c^n n^{\gamma_1} n!
  + A_2 c^{n-1} n^{\gamma_2} (n-1)!}
  {A_1 c+A_2}\, ,
\label{eq:d_n mod}
\end{eqnarray}
where the parameters $A_i$, $\gamma_i$, and $c$  are determined
from the values of the first few known coefficients $\hat{d}_n$.
We mention incidentally that the case of an alternating-sign series
has been considered in \cite{MS04} and will not be addressed here.
As regards the most general case of the series, the behavior of
$\tilde{d}^\text{model}_n$ with respect to $n$ turns out to
be the same as what is obtained in the renormalon approach (that is
usually supplied with the Naive Non-Abelization (NNA) \cite{BroGro95}),
see, e.g., \cite{BroadKa93}---except perhaps for the specific
values of the parameters which are different.
Modeling the expansion coefficients according to Eq.\
(\ref{eq:d_n mod}), leads to the following generating function $P(t)$:
\begin{eqnarray}
  P^\text{model}(t)
&\sim &
  \left[A_1 (t/c)^{1+\gamma_1}
  +  A_2 (t/c)^{\gamma_2}
  \right]\,e^{-t/c}\, .
\label{eq:d_n model}
\end{eqnarray}
Using the first already known coefficients, we can calibrate our model
in order to extract information about still higher and unknown orders,
as well as to gain information on the resumed behavior of the whole
series.
It should be emphasized that for large values of the argument $t$,
our model for $P(t)$ can become rather crude.
This, however, does not significantly change the final result of the
summation due to the convergence of the integral in Eq.\
(\ref{eq:convolution}).
For this reason, it is not necessary to know the asymptotic behavior
of $P^\text{model}(t)$ very accurately.
All said, let us now consider concrete examples to understand the
modus operandi and the benefits of this technology.

\section{Master example: Higgs-boson decay into a $\bar{b}b$ pair}
\label{sec:App.Higgs}
Our goal in this section is the calculation of the width of the
Higgs-boson decay into a $\bar{b}b$ pair, i.e.,
\begin{eqnarray}
  \Gamma_{H\to b\bar{b}}(M_{H})
=
  \Gamma_0^{b}(m_b^2)\,
  \frac{\widetilde{R}_\text{S}(M_{H}^2)}
       {3\,m_b^2}\, ,
\label{eq:Higgs.decay.rate}
\end{eqnarray}
where
$\Gamma_0^{b}(m_b^2)=3\,G_F\,M_H\,m_{b}^2/{4\sqrt{2}\pi}$,
$m_b$ and $M_H$ are the pole mass of the $b$-quark and the mass of
the Higgs boson, respectively, and
$\widetilde{R}_\text{S}(M_{H}^2)
 = \bar{m}^2_{b}(M_{H}^2)\,R_\text{S}(M_{H}^2)
$.
Our interest in this quantity derives from the fact that this
process contains all main ingredients for such high-order calculations,
discussed above, with known coefficients up to the order
$O(\alpha_s^4)$ \cite{BCK05}.
The quantity $R_\text{S}(s)$ is the discontinuity
$R_\text{S}(s)
 = \textbf{Im}\, \Pi(-s-i\epsilon)/{(2\pi\,s)}
$
of the imaginary part of the correlator $\Pi$ of the two scalar
currents $J^S_b=\bar{\Psi}_b\Psi_b$ for bottom quarks
with an on-shell mass $m_b$
coupled to the scalar Higgs boson with mass $M_H$,
and it is given by
$$
  \Pi(Q^2)
=
  (4\pi)^2 i\int dx e^{iqx}\langle 0|\;T[\;J^\text{S}_b(x)
  J^\text{S}_{b}(0)\,]\;|0\rangle\, ,
$$
where
$Q^2 = - q^2$.
Direct multi-loop calculations are usually performed in the far
Euclidean (spacelike) region for the corresponding Adler function
$D_\text{S}$ \cite{Che96,BCK05,BKM01,KK09},
where QCD perturbation theory works reliably.
Hence, we write
\begin{eqnarray}
  \frac{Q^2}{6}\frac{d}{d Q^2}\left(\frac{\Pi(Q^2)}{Q^2}\right)
=
  D_{\text{S}}(Q^2)
&=&
  \left[1+\sum_{n \geq 1} d_n\,a_s^{n}(Q^2)
  \right]\,.
\label{eq:D-s}
\end{eqnarray}

\subsection{Generating function for the scalar correlator
            and estimates for higher-order coefficients}
\label{subsec:Higgs.P(t)}
The Adler function, related to the scalar correlator
(see \cite{BCK05,BMS06}) and pertaining to the Higgs-boson decay, reads
\begin{eqnarray}
  D_\text{S}[L]
&=&
  1 + d_1\,\sum_{n=1}^{\infty}\tilde{d}_{n}\,
  \left(a_s[L]\right)^{n}\, .
\label{eq:D_S}
\end{eqnarray}
Note that here the definition of the coefficients
$\tilde{d}_{n}$ is given by
$\tilde{d}_{n}=d_{n}/d_{1}$.
The $n$-dependence of the coefficients $\tilde{d}_n$,
in accordance with expression (\ref{eq:d_n mod}),
can be simulated by the following two-parameter model
\begin{subequations}
\label{eq:Higgs.Model}
\begin{eqnarray}
\label{eq:Higgs.d_n.Model}
  \tilde{d}_n^H
=
  c^{n-1}\frac{n+\delta}{1+\delta}\,\Gamma (n)\, ,
\end{eqnarray}
a form which ensures that $\tilde{d}_1^H$
(the superscript $H$ denoting ``Higgs'')
is automatically equal to unity.
Concurrently, these coefficients can be reproduced by the following
generating function:
\begin{eqnarray}
\label{eq:Higgs.P(t).Model}
  P_H(t)
=
  \frac{(t/c)+\delta}{c\,(1+\delta)}\,e^{-{t/c}}\, .
\end{eqnarray}
\end{subequations}
To test these formulae, we use the two known coefficients
$\tilde{d}_2$ and $\tilde{d}_3$
for this process and plug them into model (\ref{eq:Higgs.d_n.Model})
aiming to predict the next two coefficients
$\tilde{d}_4^H$ and $\tilde{d}_5^H$.
The result for the next coefficient $\tilde{d}_4^H$ is shown in
the second row of Table \ref{Tab:d_n.Higgs}:
$\tilde{d}_4^H=662$.
We see that it is fairly close to the value 620 calculated recently by
Chetyrkin \textit{et al.} in \cite{BCK05}.
This procedure can be geared up to predict the next coefficient
$\tilde{d}_5$.
Indeed, taking into account the coefficient $\tilde{d}_4$ and slightly
readjusting the parameters $c$ and $\delta$ of our model
(\ref{eq:Higgs.P(t).Model}) from their previous values
$\left\{c=2.5,~\delta=-0.48\right\}$
to
$\left\{c=2.4,~\delta=-0.52\right\}$,
we find (third row in Table \ref{Tab:d_n.Higgs})
$\tilde{d}_5\approx 7826$.
This result agrees very well with the value $7782$,
derived in \cite{KaSt95} by appealing to the Principle of Minimal
Sensitivity (PMS).
Moreover, it has a reasonable agreement with our previous models
(rows 3 and 4 in Table \ref{Tab:d_n.Higgs}) and also with the
improved NNA (INNA) prediction shown in row 7, while the original
NNA prediction
(value given in row 6) fails.
We may conclude that our model calculation of the coefficient $d_5$
is congruent with the prediction following from the application of
the PMS method.
Both these approaches provide results for most coefficients $d_n$
approximately twice larger than those found by applying the
INNA approximation.

\begin{table}[h]
\caption{Perturbation Theory (PT) coefficients $\tilde{d}_n$ for
the $ D_\text{S}$ series obtained from different calculations.}
\label{Tab:d_n.Higgs}
\centerline{
\begin{tabular}{|c|c|ccccc|}\hline \hline
  & PT coefficients    &~$\tilde{d}_1\vphantom{^{|}_{|}}$~
                             &~$\tilde{d}_2$~
                                    &~$\tilde{d}_3$~
                                           &~$\tilde{d}_4$~
                                                 &~$\tilde{d}_5$~
\\ \hline \hline
1 & pQCD results from \cite{BCK05}
                       & $1\vphantom{^{|}_{|}}$
                             &~7.42~&~62.3~&~620~&~---~
\\ \hline
--& Models and estimates& \multicolumn{5}{|c|}{} \\ \hline
2 & Model (\ref{eq:Higgs.Model}) with $c=2.5,~\delta=-0.48$~
                       & $1\vphantom{^{|}_{|}}$
                             &~7.42~&~62.3~&~662~&~8615~
\\ \hline
3 & Model (\ref{eq:Higgs.Model}) with $c=2.4,~\delta=-0.52$~
                       & $1\vphantom{^{|}_{|}}$
                             &~7.50~&~61.1~&~625~&~7826~
\\ \hline
4 & ``PMS'' predictions from~\cite{ChKS97,BCK05,KaSt95}\footnote{%
 The value of the coefficient $\tilde{d}_5$ has been computed
 here with the method of \cite{KaSt95}.}~
                       &~$-\vphantom{^{|}_{|}}$~
                             &~$-$~ &~64.8~&~547~&~7782~
\\ \hline
5 & ``NNA'' predictions from~\cite{BKM01, BroadKa02}~
                       & $1\vphantom{^{|}_{|}}$
                             & 3.87~&~21.7~&~122~&~1200~
\\ \hline
6 & ``INNA'' predictions from~App.~\ref{App:INNA}\footnote{%
 We employ the upper limit of the INNA predictions---
 consult Appendix~\ref{App:INNA}.}
                       & $1\vphantom{^{|}_{|}}$
                             & 3.87~&~36~  &~251~&~4930~
\\ \hline
\hline
\end{tabular}}

\end{table}

\subsection{Higgs-boson decay width in FAPT}
 \label{subsec:Higgs.Gamma}
Within the one-loop approximation of FAPT, the quantity
$\widetilde{R}_\text{S}(M_{H}^2)$ has the following non-power series
expansion:\footnote{The appearance in the denominators of the factors
$\pi^n$ in tandem with the coefficients $\tilde{d}_n$ is a
consequence of the particular $d_n$ normalization---see
Eq.\ (\ref{eq:D_S}).}
\begin{eqnarray}
  \widetilde{R}_\text{S}^{\text{\tiny FAPT}}[L]
=
  3\,\hat{m}_{(1)}^2\,
  \left\{
         {\mathfrak A}_{\nu_{0}}^{\text{\t  iny glob}}[L]
         + d_1\,\sum_{n\geq1}\frac{\tilde{d}_{n}}{\pi^{n}}\,
         {\mathfrak A}_{n+\nu_{0}}^{\text{\tiny glob}}[L]
  \right\}\, .
\label{eq:R_S-MFAPT}
\end{eqnarray}
Here $\hat{m}_{(1)}^2$ is the renormalization-group invariant mass
satisfying the one-loop $\bar{m}^2_{b}(\mu^2)$ evolution equation
\begin{eqnarray}
  \bar{m}_{b}^2(Q^2)
=
  \hat{m}_{(1)}^2\,
  \alpha_{s}^{\nu_{0}}(Q^2)
\label{eq:m2-hat-run}
\end{eqnarray}
with $\nu_{0}=2\gamma_0/b_0(5)=1.04$, where
$\gamma_0$ is the quark-mass anomalous dimension, and the
renormalization-group invariant quantity $\hat{m}_{(1)}$ is
defined through the effective RG mass
$\overline{m}_b(\overline{m}_b^2)$.
It is worth recalling that Eq.\ (\ref{eq:R_S-MFAPT})
in the one-loop running of the coupling coincides with the result
obtained in \cite{BKM01} for $\widetilde{R}_\text{S}$ using CIPT.

The key question now is to which extent FAPT is able to reproduce with
a sufficient quality the whole sum of the series expansion of
$\widetilde{R}_\text{S}^{\text{\tiny FAPT}}[L]$
in the range $L\in[11.7,13.6]$.
This will be checked by taking recourse to the model
(\ref{eq:Higgs.Model}) in conjunction with Eq.\
(\ref{eq:Glo-MFAPT.sum.R.Glo.456}).
Note that this $L$ range corresponds to the Higgs-mass values
$M_H\in[80,180]$~GeV
with
$\Lambda^{N_f=3}_{\text{QCD}}=201$~MeV
and $\mathfrak A^{\text{\tiny glob}}_{1}(m_Z^2)=0.1226$.
In this mass range, we have $L_5<L<L_6$, so that Eq.\
(\ref{eq:Glo-MFAPT.sum.R.Glo.456}) transforms into
\begin{eqnarray}
  \Gamma_{H\to b\bar{b}}^{\text{\tiny FAPT}}[L]
=
  \Gamma_0^{b}(\hat{m}_{(1)}^2)\,
  \left\{\mathfrak A^\text{\tiny glob}_{\nu_{0}}[L]
         + \frac{d_1}{\pi}\,
           \left\langle\!\!\!\left\langle{
                             \bar{\mathfrak A}_{1+\nu_{0}}\!
                             \left[L\!+\!\lambda_5\!
                             -\!\frac{t}{\pi\beta_5}
                             \right]
                             \!+\!\Delta_{6}
                             \bar{\mathfrak A}_{1+\nu_{0}}
                             \left[\frac{t}{\pi}\right]}
           \right\rangle\!\!\!\right\rangle_{\!\!\!P_{\nu}}
  \right\}
\label{eq:R_S.Sum}
\end{eqnarray}
with $P_{\nu_{0}}(t)$ (defined via Eq.\ (\ref{eq:P.nu}))
and where we have evaluated Eq.\ (\ref{eq:Higgs.Model})
with the parameter values
$c=2.4$, $\delta=-0.52$, and $\nu_{0}=1.04$.

The all-order expression (\ref{eq:R_S.Sum}) above allows us to
determine the accuracy of the truncation procedure of the FAPT
expression
\begin{eqnarray}
  \Gamma_{H\to b\bar{b}}^{\text{\tiny FAPT}}[L;N]
&=&
  \Gamma_0^{b}(\hat{m}_{(1)}^2)\,
  \left\{
         \mathfrak A_{\nu_{0}}^{\text{\tiny glob}}[L]
         + d_1\,\sum_{n=1}^{N}
         \frac{\tilde{d}_{n}}{\pi^{n}}\,
         \mathfrak A_{n+\nu_{0}}^{\text{\tiny glob}}[L]
  \right\}
\label{eq:FAPT.trunc}
\end{eqnarray}
at order $N$ and estimate the relative errors
\begin{eqnarray}
  \Delta_N[L]
&=&
  1 - \frac{\Gamma_{H\to b\bar{b}}^{\text{\tiny FAPT}}[L;N]}
           {\Gamma_{H\to b\bar{b}}^{\text{\tiny FAPT}}[L]}\, .
\label{eq:relat.errors}
\end{eqnarray}
\begin{figure}[t!]
 \centerline{\includegraphics[width=0.47\textwidth]{
  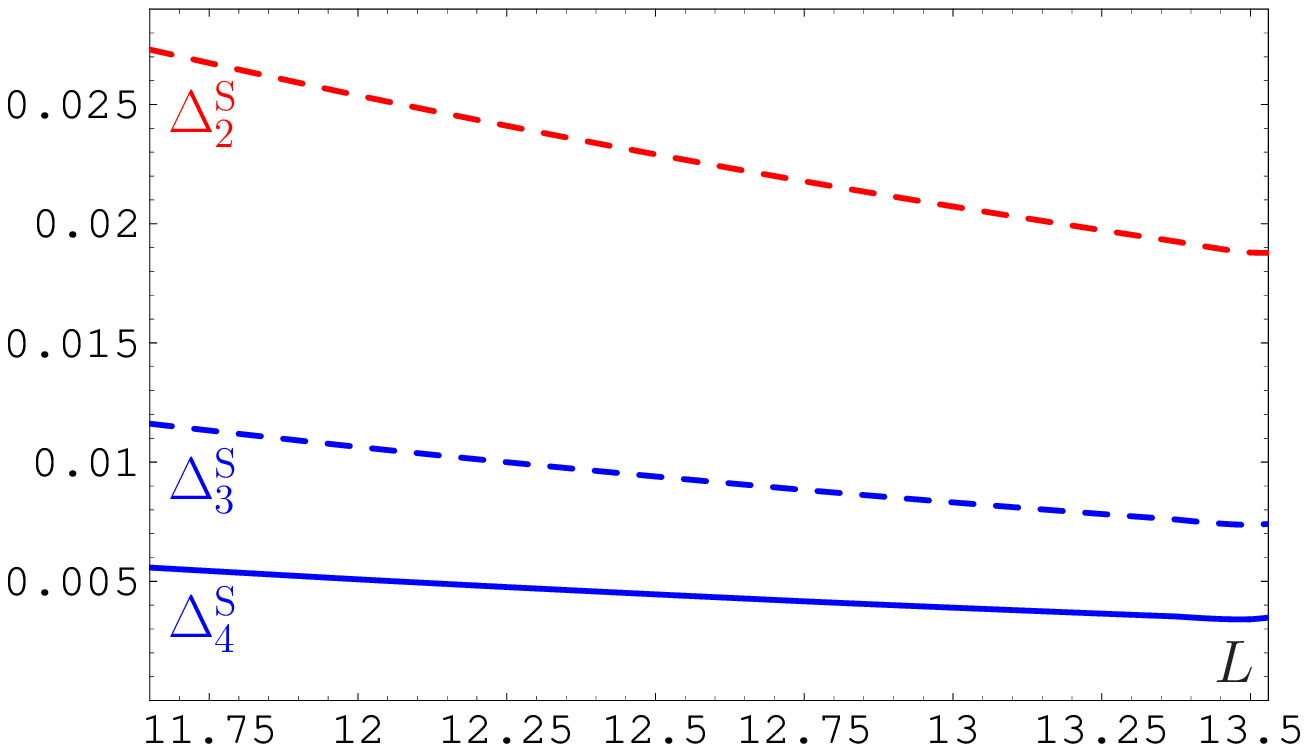}~~~\includegraphics[width=0.46\textwidth]{
  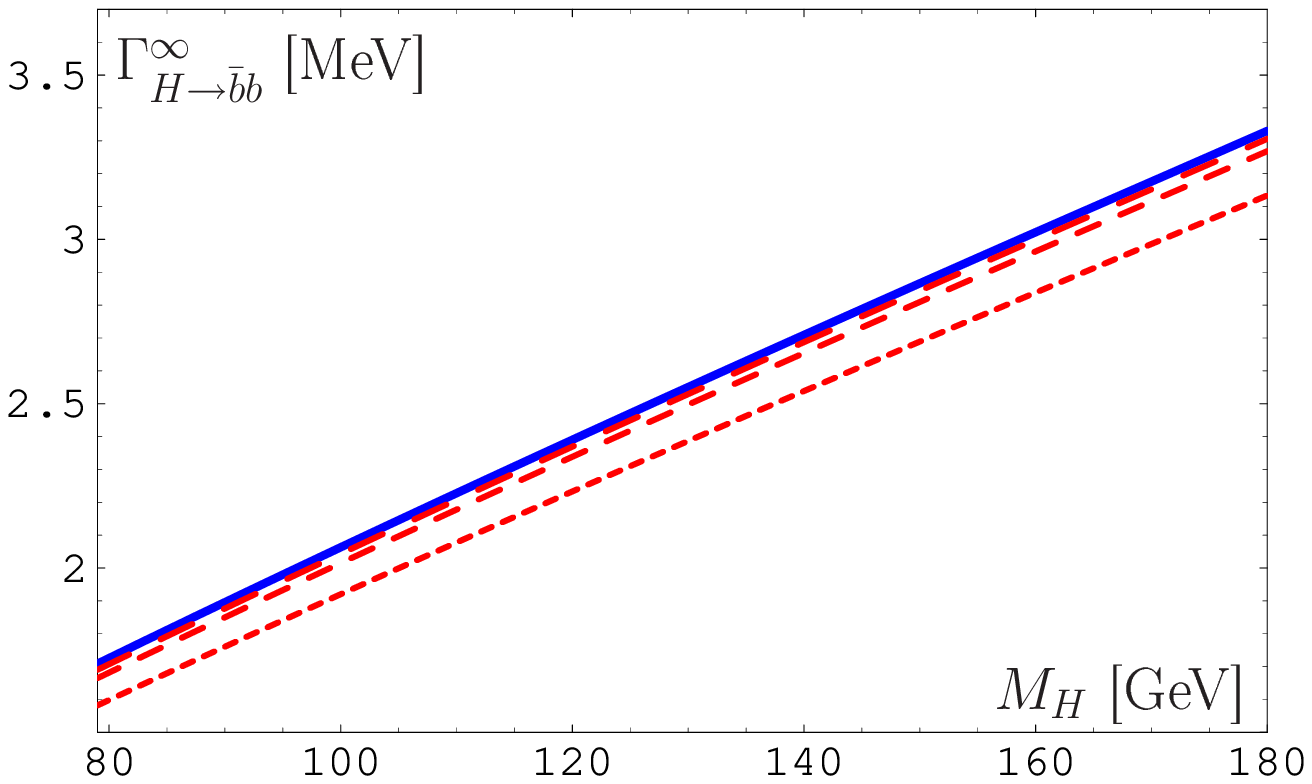}\vspace*{-1mm}}
  \caption{Left: The relative errors $\Delta_N[L]$ for $N=2, 3$, and
  $4$ of the truncated FAPT expansion, given by Eq.\
  (\ref{eq:FAPT.trunc}), with respect to the result of the resummation
  procedure, represented by Eq.\ (\ref{eq:R_S.Sum}).
  Right: The width $\Gamma_{H\to b\bar{b}}$ as a function of the
  Higgs-boson mass $M_{H}$ in the resummed (blue solid line) and
  the truncated (at the order $N$) FAPT for three different
  values of $N$ represented by dashed red lines.
  The short-dashed line corresponds to $N=1$, the dashed one to
  $N=2$, and the long-dashed line to $N=3$.
\label{fig:Higgs}}
\end{figure}
To this end, we use the values of the RG-invariant masses
$\hat{m}_{(1)}$ in the one-loop approximation, which we have
extracted from two different sources:
$\hat{m}_{b}^{(1)}=8.22$~GeV \cite{KuSt01} and
$\hat{m}_{b}^{(1)}=8.53$~GeV \cite{PeSt02},
with details being provided in Appendix \ref{app:masses.b-quark}.

\begin{figure}[h!]
 \centerline{\includegraphics[width=0.49\textwidth]{
  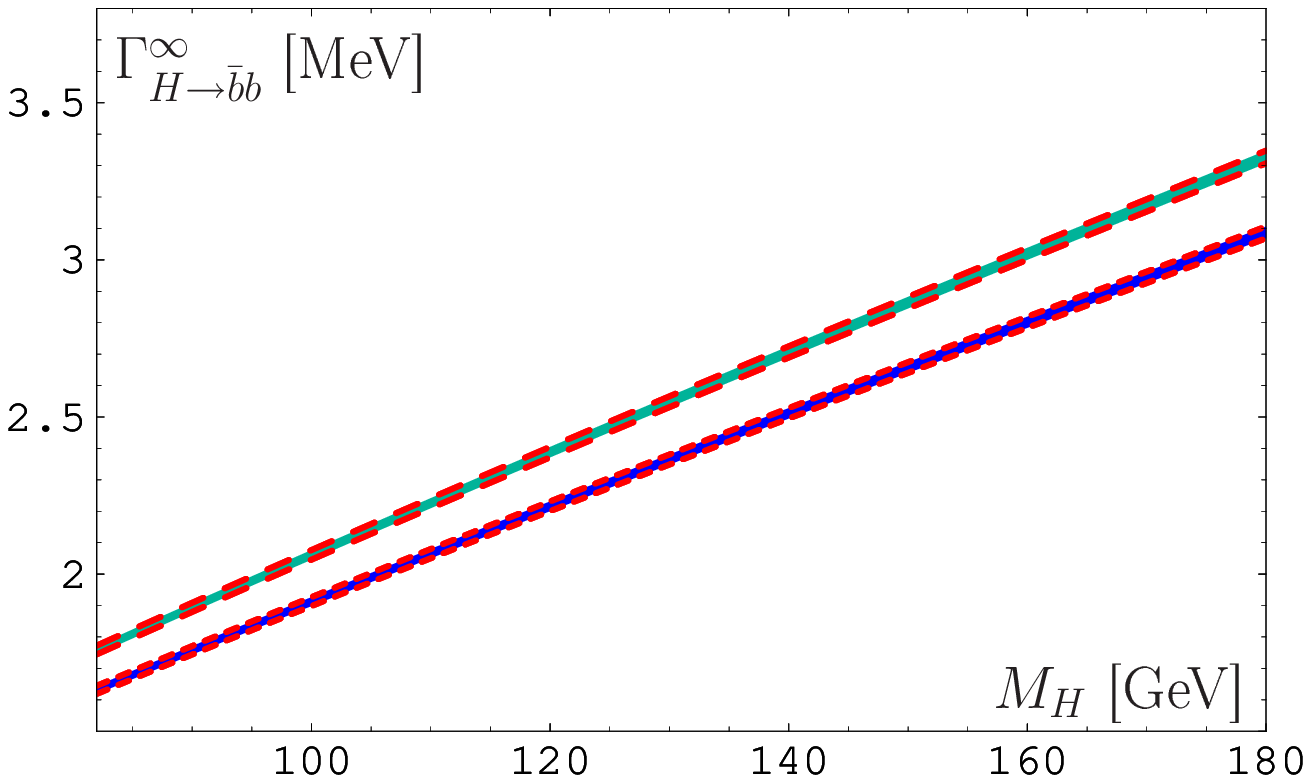}~\includegraphics[width=0.49\textwidth]{
  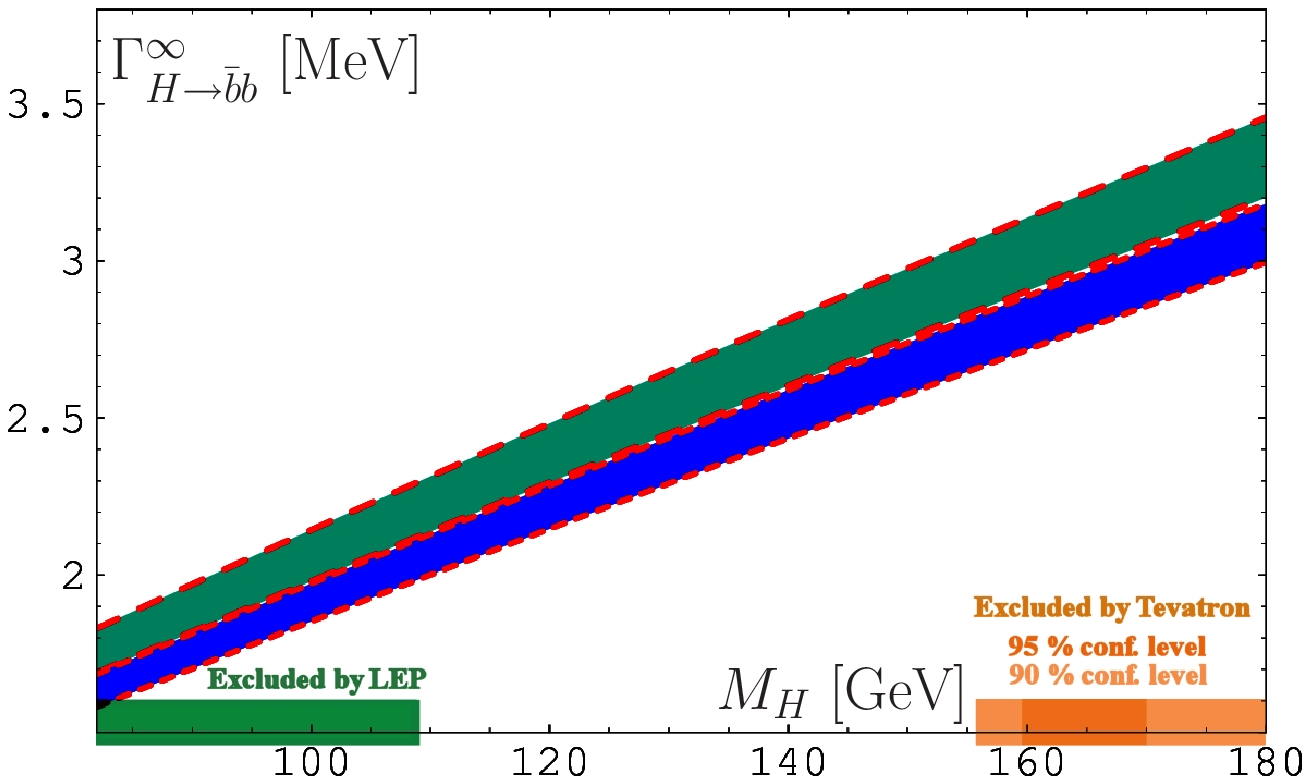}\vspace*{-1mm}}
  \caption{Left: The width $\Gamma_{H\to b\bar{b}}^{\infty}$
    as a function of the Higgs-boson mass $M_H$ in the resummed FAPT
    expansion, using different models for the generating function
    $P(t)$---see Table \ref{Tab:4}.
    Right: The width $\Gamma_{H \to b\bar{b}}^{\infty}$
    as a function of the Higgs-boson mass $M_H$ in the resummed FAPT,
    varying both the generating function $P(t)$
    in accordance with Table \ref{Tab:4} and also the mass
    $\hat{m}_{(1)}$ with $\delta \hat{m}_{(1)}=\pm0.1$~GeV.
    In both panels the upper strip corresponds to the
    Penin--Steinhauser estimate $\hat{m}_{(1)}=8.53$~GeV \cite{PeSt02},
    whereas the lower one derives from the value $\hat{m}_{(1)}=8.21$~GeV
    determined by K\"{u}hn and Steinhauser in Ref.\ \cite{KuSt01}.
    Here and also in Fig.\ \ref{fig:HiggsStrip2L} we indicated on the
    abscissa the window of the mass values of the Higgs-boson still
    accessible to experiment.
    \label{fig:HiggsStrip}}
\end{figure}

In Fig.\ \ref{fig:Higgs} we show the relative errors, given by
(\ref{eq:relat.errors}), for $N=2$, $N=3$, and
$N=4$ in the probed range of $L\in[11.7,13.6]$.
We see that already
$\Gamma_{H\to b\bar{b}}^{\text{\tiny FAPT}}[L;2]$
gives correct results to better than 2.5\%, whereas
$\Gamma_{H\to b\bar{b}}^{\text{\tiny FAPT}}[L;3]$
reaches a still better accuracy at the level of 1\%.
This means that, on practical grounds, there is no need to calculate
further corrections, because in order to be correct at the level of
1\%, it is actually sufficient to take into account only the first
three coefficients up to $\tilde{d}_3$.
This conclusion does not change if we vary the parameters of the model
$P_H(t)$.
To be more precise, varying the coefficients $\tilde{d}_i$ in a
reasonable range---in correspondence to their order, say, about 5$\%$
for $\tilde{d}_2$ up to 30$\%$ for $\tilde{d}_5$---we induce changes
of the parameter $c$ on the level of about $5\%$
which leave the main results
(and conclusions) unchanged.
By the same token, we can conclude that the quality of the convergence
of the considered series in FAPT is quite high with a tolerance of
only a few percent.
\begin{table}[h]
\caption{Perturbation Theory (PT) coefficients $\tilde{d}_n$ for
 the $ D_\text{S}$ series in the deformed versions of the model
 in comparison with the original one.}
\label{Tab:4}
\centerline{
\begin{tabular}{|c|ccccc|}\hline \hline
 PT coefficients
       & ~$\tilde{d}_1\vphantom{^{|}_{|}}$~
               &$\tilde{d}_2$
                       &$\tilde{d}_3$
                              &$\tilde{d}_4$
                                      &$\tilde{d}_5$
\\ \hline \hline
~pQCD results with $N_f=5$~\cite{BCK05}
       & 1     & 7.42~ &~62.3~&~620~  &~---~$\vphantom{^{|}_{|}}$
\\ \hline
~Normal Model (\ref{eq:Higgs.Model}) with $c=2.43,~\delta=-0.52$~
       & 1     & 7.50~ &~61.1~&~625~  &~7826~$\vphantom{^{|}_{|}}$
\\ \hline
~Enhanced Deformation (\ref{eq:Higgs.Model}) with $c=2.62,\delta=-0.50$
       & 1     & 7.85 & 68.5 & 752 & 10120$\vphantom{^{|}_{|}}$
\\ \hline
~Reduced Deformation (\ref{eq:Higgs.Model}) with $c=2.25,\delta=-0.51$~
       & 1     & 6.89 & 52.0 & 492 & 5707$\vphantom{^{|}_{|}}$
\\ \hline \hline
\end{tabular}}
\end{table}
Let us now expand our statements about the uncertainties of our
results with regard to the model generating function $P(t)$.
To this end, we deform our original model with $c=2.43$ and
$\delta=-0.52$ in order to enhance or reduce the magnitude of
the last known coefficient $\tilde{d}_4$ and the value of the still
uncalculated coefficient $\tilde{d}_5$
(for details see Table \ref{Tab:4}).
The results of this variation are shown in the left panel of Fig.\
\ref{fig:HiggsStrip} in the form of a strip, the upper boundary of
which is formed by the enhanced version of the $P(t)$ model, whereas
the lower boundary of the strip corresponds to the ``reduced'' version
of the model.
We see that the uncertainty caused by this deformation is less than
0.5\%.

Now we want to discuss what changes are induced when one applies for
the same kind of analysis the FAPT resummation approach at the two-loop
order.
In that case, there is a technical complication:
The evolution factors are not simple powers $a^{\nu}[L]$,
but more involved expressions like
$a^{\nu_0}[L]\left[1+c_1\,a\right]^{\nu_1}$,
as one sees by comparing with Eq.\ (\ref{eq:mass.Ev.2L}).
For this reason, the result of the resummation is more complicated,
finally amounting to Eq.\ (\ref{eq:Resum.2L.Evo}).

\begin{figure}[t!]
 \centerline{\includegraphics[width=0.49\textwidth]{
  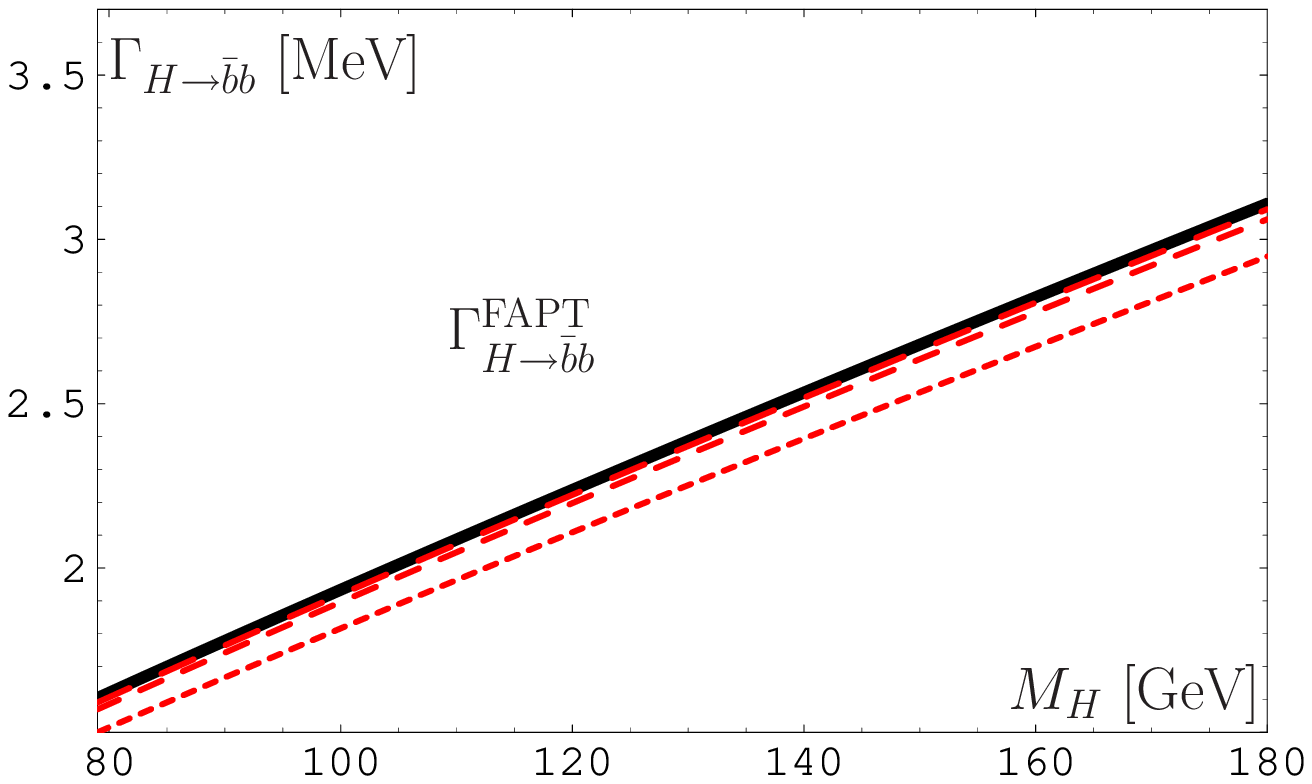}~\includegraphics[width=0.49\textwidth]{
  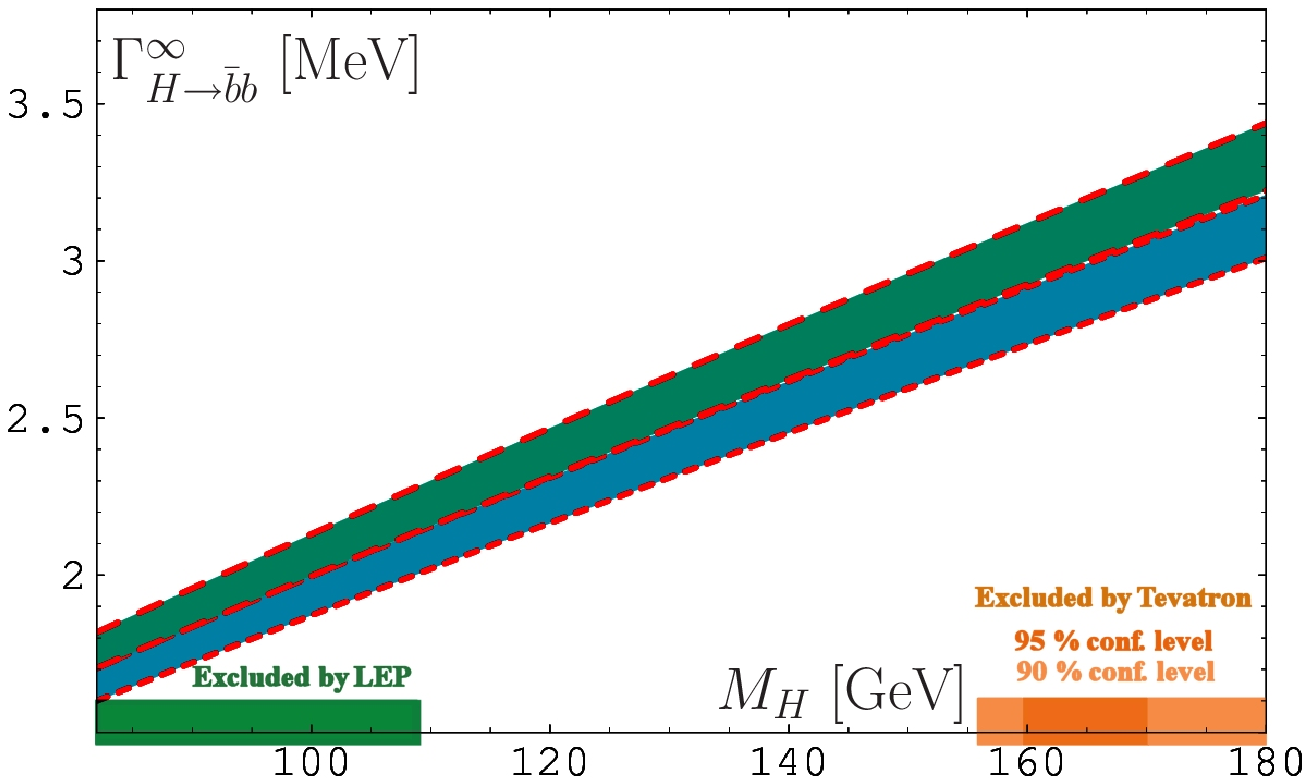}\vspace*{-1mm}}
  \caption{Left panel:~The two-loop width $\Gamma_{H\to b\bar{b}}$ as a
    function of the Higgs boson mass $M_{H}$ in the resummed (black
    solid line) and the truncated (at the order $N$) FAPT.
    Here, the short-dashed line corresponds to $N=1$, the dashed one to
    $N=2$, and the long-dashed line to $N=3$.
    Right panel:~The two-loop width $\Gamma_{H \to b\bar{b}}^{\infty}$
    as a function of the Higgs-boson mass $M_H$ in the resummed FAPT
    is shown (lower strip), varying the mass in the interval
    $\hat{m}_{(2)}=8.22\pm0.13$~GeV according to the Penin--Steinhauser
    estimate
    $\overline{m}_b(\overline{m}_b^2)=4.35\pm0.07$~GeV~\cite{PeSt02}.
    The upper strip shows the corresponding one-loop result.
    The window of the mass values of the Higgs-boson which is
    still accessible to experiment is explicitly indicated.
    \label{fig:HiggsStrip2L}}
\end{figure}

Here we resort only to graphical illustrations of our results.
In the left panel of Fig.\ \ref{fig:HiggsStrip2L}, we discuss the
convergence properties of the decay widths, truncated at the order $N$,
relative to the resummed two-loop result
$\Gamma_{H \to b\bar{b}}^{\infty}$.
From this, we infer that our conclusions drawn from the one-loop
analysis remain valid.
Indeed, $\Gamma_{H\to b\bar{b}}^{\text{\tiny FAPT}}[L;2]$
is correct to better than 2\%,
whereas $\Gamma_{H\to b\bar{b}}^{\text{\tiny FAPT}}[L;3]$
reaches an even higher precision level of the order of 0.7\%.

In the right panel of Fig.\ \ref{fig:HiggsStrip2L}, we show the
results for the decay width
$\Gamma_{H \to b\bar{b}}^{\infty}(M_H)$
in the resummed two-loop FAPT, in the window of the Higgs mass
allowed by existing experiments---LEP and Tevatron \cite{Tevatron2010}.
Comparing this outcome with the one-loop result---upper strip in the
same panel of this figure---reveals a 5\% reduction of the two-loop
estimate.
This reduction consists of two parts:
one part ($\approx+7\%$) is due to the difference in the mass $\hat{m}$
in both approximations, while the other ($\approx-2\%$) comes from the
difference in the values of $R_{S}(M_H)$ in the one-loop and the
two-loop approximations.

In our predictions we have considered two different options for the
values of the RG-effective $b$-quark mass,
$\overline{m}_b(\overline{m}_b^2)$,
which we have taken from two independent analyses.
One value originates from Ref.\ \cite{PeSt02}
($\overline{m}_b(\overline{m}_b^2)=4.35\pm0.07$~GeV),
while the other was derived in Ref.\ \cite{KuSt01}
yielding a somewhat deviating result, notably,
$\overline{m}_b(\overline{m}_b^2)=4.19\pm0.05$~GeV.
In Fig.\ \ref{fig:HiggsStrip2L} we show all two-loop quantities
adopting for the RG-effective $b$-quark mass the result
obtained by Penin and Steinhauser in Ref.\ \cite{PeSt02}.
The relative difference between these two choices is of the order of
4\%, so that squared masses differ by 7\%.
This means that the corresponding curves for the K\"{u}hn--Steinhauser
value of $\overline{m}_b(\overline{m}_b^2)$ \cite{KuSt01}
can be actually obtained from those shown by downsizing them
with an overall reduction factor of 7\%.
From this we conclude that the real theoretical uncertainty of the
Higgs-boson width in this decay channel is de facto determined
by the upper boundary of the Penin--Steinhauser estimate \cite{PeSt02},
with the lower boundary of the K\"{u}hn--Steinhauser
\cite{KuSt01} being in the range of $6$\% (6=(5+7)/2).

\section{Adler function of the vector correlator and
$\bm{R_{\lowercase{e^+e^-\to\,\textbf{hadrons}}}}$}
\label{sec:App.Adler}
So far, we have discussed only the Adler function related to the scalar
correlator.
But this sort of considerations can be applied to the vector correlator
as well.
To be specific, we are interested in modeling the generating function
of the perturbative coefficients $\tilde{d}_n$
(see the first row in Table \ref{Tab:r_n.d_k})
of the Adler function of the vector correlator
(labeled below by the symbol V)
\cite{BCK08,BCK10}
\begin{eqnarray}
  D_\text{V}[L]
&=& d_0
  + d_1\,\sum_{n=1}^{\infty}\tilde{d}_{n}\,
  \left(\frac{\alpha_s[L]}{\pi}
  \right)^{n}\, .
\label{eq:D_V}
\end{eqnarray}

To account for the $n$-dependence of the coefficients $\tilde{d}_n$,
in accordance with the asymptotic model of (\ref{eq:d_n.BLM}), we
write
\begin{subequations}
\label{eq:Vector.Model}
\begin{eqnarray}
\label{eq:Vector.d_n.Model}
  \tilde{d}_n^\text{V}
=
  c^{n-1}\,
  \frac{\delta^{n+1}-n}
  {\delta^2-1}\,\Gamma(n)\, ,
\end{eqnarray}
which can be derived from the generating function
\begin{eqnarray}
\label{eq:Vector.P(t).Model}
  P_\text{V}(t)
&=&
  \frac{\delta\,e^{-t/c\delta} - (t/c)\,e^{-t/c}}
  {c\left(\delta ^2-1\right)}\, .
\end{eqnarray}
\end{subequations}
Our predictions, obtained with this generating function by fitting
the two known coefficients $\tilde{d}_2$ and $\tilde{d}_3$ and using
the model (\ref{eq:Vector.Model}), have been included in
Table \ref{Tab:d_n.Adler}.\footnote{
Note that $\tilde{d}_1^\text{V}$ is automatically equal to unity.}
We observe a good agreement between our estimate
$\tilde{d}_4^\text{V}=27.1$
and the value 27.4, calculated recently by Chetyrkin \textit{et al.}
in Ref.\ \cite{BCK08,BCK10}.
Would we use instead a fitting procedure, which would take into
account the fourth-order coefficient $\tilde{d}_4$ in order to
predict $\tilde{d}_5$, we would have to readjust the model parameters
in (\ref{eq:Vector.Model}) to the new values
$\left\{c=3.5548,~\delta=1.32448\right\}$
$\to$
$\left\{c=3.5526,~\delta=1.32453\right\}$.
These findings provide evident support for our model evaluation,
and we may expect that our procedure will work in other cases as well.

In order to explore to what extent the exact knowledge of the
higher-order coefficients $\tilde{d}_n$ is important, we employed our
model (\ref{eq:Vector.Model}) with different values of the parameters
$c$ and $\delta$:
$c=3.63$ and $\delta=1.3231$.
These values are, roughly speaking, tantamount to replacing the exact
value of the coefficient $\tilde{d}_4=27.4$ by something approximately
equal to the NNA prediction obtained in \cite{BroadKa93,BroadKa02}.
The difference between the analytic sums of the two models in the
region corresponding to $N_f=4$ is indeed very small, reaching just
a mere $0.2\%$.

\begin{table}[ht]
\caption{Coefficients $\tilde{d}_n$ for the Adler-function series with
$N_f=4$.
The numbers in the square brackets denote the lower and the upper
limits of the INNA estimates.}
\label{Tab:d_n.Adler}
\centerline{
\begin{tabular}{|c|c|ccccc|}\hline \hline
 & PT coefficients     &~$\tilde{d}_1\vphantom{^{|}_{|}}$~
                             &~$\tilde{d}_2$~
                                    &~$\tilde{d}_3$~
                                           &~$\tilde{d}_4$~
                                                  &~$\tilde{d}_5$~
\\ \hline \hline
1 &~pQCD results with $N_f=4$ \cite{BCK08,BCK10}~
                       & $1\vphantom{^{|}_{|}}$
                             &~1.52~&~2.59~&~27.4~&~---~
\\ \hline
  &~Models and estimates~
                       & \multicolumn{5}{|c|}{} \\ \hline
2 &~Model (\ref{eq:Vector.Model}) with $c=3.555,~\delta=1.3245$~
                       & $1\vphantom{^{|}_{|}}$
                             &~1.52~&~2.59~&~27.1~&~2024~
\\ \hline
3 &~Model (\ref{eq:Vector.Model}) with $c=3.553,~\delta=1.3245$~
                       & $1\vphantom{^{|}_{|}}$
                             &~1.52~&~2.60~&~27.3~&~2025~
\\ \hline
4 &~Model (\ref{eq:Vector.Model}) with $c=3.630,~\delta=1.3231$~
                       & $1\vphantom{^{|}_{|}}$
                             &~1.53~&~2.26~&~20.7~&~2020~
\\ \hline
5 &~``NNA'' prediction of~\cite{BroadKa02, BroadKa93}
                       & $1\vphantom{^{|}_{|}}$
                             &~1.44~&~13.47~&~19.7~&~579~
\\ \hline
6 &~``INNA'' prediction of~App.~\ref{App:INNA}~
                       & $1\vphantom{^{|}_{|}}$
                             &~1.44~&~$[3.5,9.6]$~
                                            &~$[20.4,48.1]$~
                                                   &~$[674,2786]$~
\\ \hline
7 & ``FAC'' prediction of~\cite{KaSt95,BCK02}
                       & $\vphantom{^{|}_{|}}$
                             &      &       & ~8.4 $\pm 18$~
                                                   &~152~
\\ \hline \hline
\end{tabular}}
\end{table}

\begin{figure}[h!]
 \centerline{\includegraphics[width=0.5\textwidth]{
  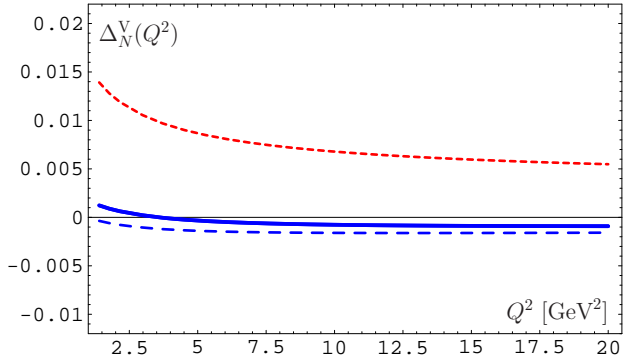}\vspace*{-1mm}}
   \caption{The relative errors $\Delta^\text{V}_N(Q^2)$ evaluated for
    different values of $N$:
    $N=1$ (short-dashed red line),
    $N=2$ (solid blue line), and
    $N=3$ (dashed blue line) of the truncated APT given by
    Eq.\ (\ref{eq:R_V.APT.trunc}), in comparison with the exact
    result of the resummation procedure represented by
    Eq.\ (\ref{eq:sum.R.Glo.456}).
   \label{fig:Adler}}
\end{figure}

Then, we have
\begin{eqnarray}
\label{eq:R_V.APT.exact}
  D_\text{V}^{\text{APT}}[L]
&=&
  1 + d_1\,\sum_{n\geq1}
             \frac{\tilde{d}^\text{V}_{n}}{\pi^{n}}\,
              \mathcal A_{n}^{\text{\tiny glob}}[L]
\, , \\
\label{eq:R_V.APT.trunc}
  D_\text{V}^{\text{APT}}[L;N]
&=&
  1 + d_1\,\sum_{n=1}^{N}
      \frac{\tilde{d}^\text{V}_{n}}{\pi^{n}}\,
      \mathcal A_{n}^{\text{\tiny glob}}[L]\, ,
\end{eqnarray}
while the global-APT resummation result for
$D_\text{V}^{\text{APT}}[L]$ is given in
Appendix \ref{app:Sum.FAPT.Global} by
Eq.\ (\ref{eq:sum.D.Glo.456}).
The relative errors
\begin{eqnarray}
  \Delta_N^\text{V}(Q^2)
&=&
   1 - \frac{D_\text{V}^{\text{APT}}[\ln(Q^2/\Lambda^2);N]}
            {D_\text{V}^{\text{APT}}[\ln(Q^2/\Lambda^2)]}\, ,
\label{eq:relat.errors.D.V.APT}
\end{eqnarray}
evaluated in the range
$Q^2\in[2,20]$~GeV$^{2}$ for three values of the expansion
$N=1, 2, 3$,
are displayed in Fig.\ \ref{fig:Adler}.
We observe from this figure that already
$D_\text{V}^{\text{APT}}[L;1]$
provides an accuracy in the vicinity of 1\%, whereas
$D_\text{V}^{\text{\tiny APT}}[L;2]$
is smaller then 0.1\% in the interval $Q^2=1-20$~GeV$^2$.
This means that there is no real need to calculate further corrections.
Staying at the level of being correct to a better accuracy than 1\%,
it is virtually enough to take into account only the terms up to $d_2$.
This conclusion is quite robust against the variation of the
parameters of the model $P_\text{V}(t)$.
The main outcome here may look somewhat surprising:
In fact, the best order of truncation of the FAPT series in the region
$Q^2=2-20$~GeV$^2$
is reached by employing the N$^2$LO approximation,
i.e., by keeping just the $d_2$-term.

\section{Conclusions}
\label{sec:conclusions}

In this work we have given considered in detail relations among popular
perturbative approaches: FOPT, CIPT, and FAPT.
We proved that in the Minkowski region both CIPT and FAPT produce
for the $R$-ratios which are related to the corresponding Adler
functions---see
Eqs.\ (\ref{eq:theorem-1})--(\ref{eq:tho-contours})---the
same results.
These results \textit{do not coincide}---for any fixed order of the
perturbative expansion---with those obtained with FOPT.

We also considered in detail the resummation properties
of non-power-series expansions within FAPT.
In particular, we have given analysis of the Adler function
of a scalar, $D_S$, and a vector, $D_V$, correlator presenting
results at the two-loop running of the coupling.
Using a particular generating function for the coefficients of the
perturbative expansion, which embodies information about their
asymptotic behavior, we derived results for the whole series by
resumming it.
We used this key feature of the non-power-series expansion within FAPT
in order to reduce the theoretical uncertainties in obtaining
estimates for crucial observables,
like the decay width of the Higgs boson into a $b\bar b$ pair,
relevant for the Higgs search at the Tevatron and the LHC
(Sec.\ \ref{sec:App.Higgs}).

Employing an appropriate generating function, we estimated the values
of the coefficients
$\tilde{d}_{4}^{H}$ and $\tilde{d}_{5}^{H}$
and found that they are close to those computed by other groups using
different perturbative methods.
Moreover, we were able to resum the whole series and extract reliable
predictions for the width $\Gamma_{H\to b\bar{b}}$
in terms of the Higgs-boson mass in comparison with analogous
estimates at fixed orders
$N=1, 2, 3$ of the perturbative expansion.
This allowed us to gauge the accuracy of the truncation procedure of
the FAPT expansion, estimating the relative errors for $N=2, 3, 4$.
We found that the convergence of the FAPT non-power-series expansion
involves uncertainties at the level of only a few percent.
A cautious conclusion from this is that the reached accuracy
of the order of 1\% at the truncation level of N$^3$LO is comparable
with, or even slightly better than, the 2\% uncertainty involved
in the $\overline{m}_b(\overline{m}_b^2)$ estimates.
For this reason, there is no real need to take into account
higher-order corrections.

Our Higgs-decay predictions are given in graphical form in Figs.\
\ref{fig:Higgs} (one-loop-order running), \ref{fig:HiggsStrip}
(one-loop-order running), and \ref{fig:HiggsStrip2L}
(two-loop-order running).
In Fig.\ \ref{fig:HiggsStrip} we presented our one-loop resummed
FAPT results for two different estimates of the mass scales:
$\hat{m}_{(1)}=8.53$~GeV \cite{PeSt02} (upper strip) and
$\hat{m}_{(1)}=8.21$~GeV \cite{KuSt01} (lower strip).
The graphical representation of the two-loop decay width in the
resummed FAPT is displayed in Fig.\ \ref{fig:HiggsStrip2L}---lower
strip---in comparison with the one-loop result (upper strip).
In this calculation we varied the mass in the interval
$\hat{m}_{(2)}=8.22\pm0.13$~GeV,
implementing the Penin--Steinhauser estimate
$\overline{m}_b(\overline{m}_b^2)=4.35\pm0.07$~GeV~\cite{PeSt02}.
The corresponding predictions for the other option, offered by
the K\"{u}hn--Steinhauser estimate
$\overline{m}_b(\overline{m}_b^2)$~\cite{KuSt01}, can be obtained
by reducing the previous results by an overall factor of about 7\%.
On the experimental side, one should keep in mind that the
decay mode $H\to b\bar{b}$ is very challenging, but due to be
measured by the ATLAS Collaboration at LHC \cite{ATLAS2010}.

In Sec.\ \ref{sec:App.Adler} we turned our attention to the Adler
function, related to a vector correlator, in order to perform
the FAPT resummation in the same manner as we did for the
Higgs-boson-decay width.
We used the generating function (\ref{eq:Vector.P(t).Model})
which depends on two parameters.
To validate the robustness of our predictions, we varied the values
of these parameters and found that this variation exerts only a small
effect on the predicted value of $\tilde{d}_5$ and on the
resummation result.
It turns out that for a fixed number of flavors $N_f=4$, the obtained
results are within the limits set by the INNA method, being, however,
incompatible with both the NNA prediction of \cite{BroadKa02,BroadKa93}
and also the FAC one \cite{KaSt95,BCK02}.
We also estimated the influence of heavy-quark thresholds crossing
in Appendix \ref{app:Sum.FAPT.Global}.

Bottom line: We provided evidence in terms of two concrete examples
that FAPT can provide accurate and robust estimates for relevant
observables that are otherwise inaccessible by FOPT or CIPT.
Moreover, using the suggested resummation approach within FAPT,
we could optimize the truncation of the perturbative
non-power-series expansion, thus minimizing the truncation uncertainty.

\acknowledgments
We would like to thank  Konstantin Chetyrkin, Friedrich Jegerlehner,
Andrei Kataev, Viktor Kim, Alexei Pivovarov, and Dmitry Shirkov
for stimulating discussions and useful remarks.
N.G.S. is grateful to BLTP at JINR for support, where most of this work
was carried out.
A.B. and S.M. acknowledge financial support from Nikolay Rybakov.
This work received partially support from the Heisenberg--Landau
Program under Grants 2008, 2009, and 2010, the Russian Foundation for
Fundamental Research (Grants No.\ 07-02-91557, No.\ 08-01-00686, and
No.\ 09-02-01149), and the BRFBR--JINR Cooperation Program, contract
No.\ F06D-002.

\begin{appendix}
\appendix
\section{Relations between $R$ and $D$}
 \label{App:A}
In this Appendix we derive expression (\ref{eq:one-loop.r_n}) basing
our reasoning on the useful relations between $R$ (LHS) and $D$ (RHS)
\begin{subequations}
\label{eq:A.1}
\begin{eqnarray}
  \int_0^\infty{ds\over(s+Q^2)}\left({\mu^2\over s}\right)^\delta
&=&
  {\pi\over\sin(\pi\delta)}
  \left({\mu^2\over Q^2}\right)^\delta\,;
\label{eq:pi1} \\
  Q^2\int_0^\infty{ds\over(s+Q^2)^2}\left({\mu^2\over s}\right)^\delta
&=&
  {\pi\delta\over\sin(\pi\delta)}
  \left({\mu^2\over Q^2}\right)^\delta\, ,
\label{eq:pi2}
\end{eqnarray}
\end{subequations}
worked out in \cite{BKM01}.
Equation (\ref{eq:pi2}) can be rewritten in the form of a dispersion
relation between two specific quantities $\tilde{R}$ and $\tilde{D}$
to read
\begin{eqnarray}
 \hat{D}\left[\tilde{R}\right]
\equiv
  Q^2\int_0^\infty{ds\over(s+Q^2)^2}\tilde{R}
=
  Q^2\int_0^\infty{ds\over(s+Q^2)^2}
  \left[{\sin(\pi\delta)\over\pi\delta}
        \left(\frac{\mu^2}{s}\right)^\delta
  \right]
=
  \left({\mu^2\over Q^2}\right)^\delta\,=\tilde{D}\, .~~~
\label{eq:pi3}
\end{eqnarray}
From this equation we deduce the following correspondence between the
terms $\tilde{R}$ (LHS) and $\tilde{D}$ (RHS):
\begin{eqnarray}
\label{eq:A.pi3}
  \tilde{R}
=
  {\sin(\pi\delta)\over\pi\delta}\left(\frac{\mu^2}{s}\right)^\delta
\Longleftrightarrow
  \left(\frac{\mu^2}{Q^2}\right)^\delta
= \tilde{D}\, .
\end{eqnarray}

The powers of $\ln(Q^2/\mu^2)$ in $D(Q^2,\mu^2)$ generate, by means of
$\hat{R}\left[D\right]$ in Eq.\ (\ref{eq:R-D-operation}), the
$\pi^2$-terms in $R(s,\mu^2)$.
To demonstrate how this happens, we take $2k$ times derivatives with
respect to the variable $\pi\delta$ on both sides of Eq.\
(\ref{eq:A.pi3}) and set $\delta=0$.
Then, we obtain a new relation between $\tilde{D}$ and $\tilde{R}$,
given by
\begin{eqnarray}
\label{eq:A.pi4}
  \tilde{D} \to \left(\ln(Q^2/\mu^2) \right)^{2 k}
\Longleftrightarrow
  (-1)^k \frac{(\pi^2)^{ k}}{2k+1}+ \ldots \, ,
\end{eqnarray}
where the ellipsis on the RHS indicates that the terms with the
powers of $\ln(s/\mu^2)$ in $R$ should, ultimately, disappear upon
setting $s=\mu^2$, because only the even powers will survive.
This can be traced back to the fact that the
$\left(\ln(Q^2/\mu^2) \right)^{m}$ terms in $D$ originate from the
expansion of the RHS of equation
$$ d_n a^n_s(Q^2)
 = d_n a^n_s(\mu^2)
  \left(1+a_s(\mu^2)\beta_0 \ln(Q^2/\mu^2)\right)^{-n}
$$
that leads to the expression
\begin{eqnarray}
\label{eq:A.exp}
  d_n\,a^n_s(Q^2)
= d_n\,a_s^n \sum_{m=0}^{}
  \frac{(n-1+m)!}{(n-1)!~(m)!}
  (a_s\beta_0 \ln(Q^2/\mu^2) )^{m}\, .
\end{eqnarray}
Finally, substituting relation (\ref{eq:A.pi4}) into the expansion
entering Eq.\ (\ref{eq:A.exp}), one arrives at
Eq.\ (\ref{eq:one-loop.r_n})
\begin{eqnarray}
  d_n\,a^n_s
\to
  d_n\,a_s^n \sum_{k=0}^{}
  \frac{(n-1+2 k)!}{(n-1)!~(2 k)!}
  (-1)^k
  \frac{(a_s\beta_0 \pi )^{2k}}{2k+1}\, .
\end{eqnarray}

\section{Complete global expressions for $D$ and $R$}
 \label{app:Sum.FAPT.Global}
In this appendix, we derive explicit expressions for
$R^\text{\tiny glob}[L]$ and
${D}^\text{\tiny glob}[L]$
valid in the global scheme of APT.
The spectral density
$\rho^\text{\tiny glob}_n(\sigma)$
is related to the spectral density
$\bar{\rho}_n(\sigma;N_f)$~\cite{SM94,SS96,Mag99}:
\begin{eqnarray}
  \rho^\text{\tiny glob}_n(\sigma)
= \theta\left[\sigma\leq m^2_{4}\right]\,
  \bar{\rho}_n(\sigma;3)
  + \sum\limits_{f=4}^{6}
  \theta\left[m^2_{f}<\sigma\leq m^2_{f+1}\right]\,
  \bar{\rho}_n(\sigma;N_f)\, ,
\label{eq:rho_n.Glo}
\end{eqnarray}
where---in order to make our formulae more compact---we wrote
$m_4=m_c$, $m_5=m_b$, $m_6=m_t$, and $m_7=+\infty$.

Then, as was shown in \cite{BM08,AB08gfapt}, we have for
\begin{subequations}
\begin{eqnarray}
  R^\text{\tiny FAPT}[L]
=
  d_0
  \!&\!+\!&\! d_1\sum\limits_{f=3}^{6}
  \theta\left(L_{f}\!\leq\!L\!<\!L_{f+1}\right)
  \left\langle\!\!\!\left\langle%
  {\bar{\mathfrak A}_{1}\!\Big[L\!+\!\lambda_f\!-\!\frac{t}{\beta_f};f
                           \Big]
  }\right\rangle\!\!\!\right\rangle_{P}
\nonumber\\
  \!&\!+\!&\! d_1\sum\limits_{f=3}^{5}
  \theta\left(L_{f}\!\leq\!L\!<\!L_{f+1}\right)
  \sum\limits_{k=f+1}^{6}
  \left\langle\!\left\langle{\Delta_{k}\bar{\mathfrak A}_{1}[t]}
  \right\rangle\!\right\rangle_{P}\,;~~~
\label{eq:sum.R.Glo.456}\\
 {D}^\text{\tiny FAPT}[L]
=
  d_0 \!&\!+\!&\!
  d_1\,\sum\limits_{f=3}^{6}
  \left\langle\!\!\!\left\langle { \int_{L_{f}}^{L_{f+1}}\!
  \frac{\bar{\rho}_{1}\left[L_{\sigma}+\lambda_f;N_f\right]
  \,dL_\sigma}
  {1+e^{L-L_{\sigma}-t/\beta_f}} }
  \right\rangle\!\!\!\right\rangle_{P}
\nonumber\\
\!&\!+\!&\!
     d_1\,\sum\limits_{f=4}^{6}
     \left\langle\!\left\langle{
     \Delta_{f}[L,t]}
     \right\rangle\!\right\rangle_{P}\, ,
\label{eq:sum.D.Glo.456}
\end{eqnarray}
\end{subequations}
where
$\lambda_f\equiv\ln(\Lambda_3^2/\Lambda_f^2)$
describes the shift of the logarithmic argument due to the change of
the QCD scale parameter
$\Lambda_3\to\Lambda_f$,
(with $L_{f}\equiv\ln(m_f^2/\Lambda_3^2)$),
$L_3=-\infty$, and $L_7=+\infty$.
The second term in Eq.\ (\ref{eq:sum.R.Glo.456}) presents a natural
generalization of the fixed-$N_f$ formula (\ref{eq:APT.Sum})
for the case of using different QCD scales $\Lambda_f$ for
each fixed-$N_f$ interval.
The last term in Eq.\ (\ref{eq:sum.D.Glo.456}) appears because of the
continuation of
$R^\text{\tiny glob}[L]$
at the heavy-quark thresholds:
\begin{eqnarray}
 \Delta_{f}\bar{\mathfrak A}_{1}[t]
\equiv
 \bar{\mathfrak A}_{1}\!
 \Big[L_{f}+\lambda_{f}-\frac{t}{\beta_f};f\Big]
 - \bar{\mathfrak A}_{1}\!
 \Big[L_{f}+\lambda_{f-1}-\frac{t}{\beta_{f-1}};f-1\Big]\, .
\label{eq:Glo-MFAPT.Delta.1+nu}
\end{eqnarray}
In the Euclidean domain, the corrections
$
 \left\langle\!\left\langle{
 \Delta_{f}[L,t]}\right\rangle\!\right\rangle_{P}
$
to the naive expectation formula are defined by
\begin{eqnarray}
  \Delta_{f}[L,t]
&\equiv&
  \int_{0}^{1}\!
  \frac{\bar{\rho}_{1}\left[L_f+\lambda_f-tx/\beta_f;N_f\right]\,t}
  {\beta_f\,\left[1+e^{L-L_f-t\bar{x}/\beta_f}\right]}\,dx\nonumber\\
  &-& \int_{0}^{1}\!
  \frac{\bar{\rho}_{1}
  \left[L_f+\lambda_{f-1}-tx/\beta_{f-1};N_{f-1}\right]\,t}
  {\beta_{f-1}\,\left[1+e^{L-L_f-t\bar{x}/\beta_{f-1}}\right]}\,dx\,
.~~~
\label{eq:Delta.f.Eucl}
\end{eqnarray}
In contrast to the Minkowski case, they explicitly depend on $L$.

Consider now the extension of these summation techniques to global
FAPT, i.e., when one takes into account heavy-quark thresholds.
To be more precise, we will deal with the summation of the
following series:
\begin{eqnarray}
  {D_\nu^\text{\tiny FAPT}[L]
\choose
  R_\nu^\text{\tiny FAPT}[L]
  }
=
  d_0
  {\mathcal A^\text{\tiny glob}_{\nu}[L]
\choose
  \mathfrak A^\text{\tiny glob}_{\nu}[L]
  }
  + d_1\,\sum_{n=1}^{\infty}\tilde{d}_{n}\,
  {\mathcal A^\text{\tiny glob}_{n+\nu}[L]
\choose
  \mathfrak A^\text{\tiny glob}_{n+\nu}[L]
   }\, .
\label{eq:Sum.DR.Global}
\end{eqnarray}
Note that due to the different relative normalization of
$\mathcal A_{n+\nu}[L]$ ($\mathfrak A_{n+\nu}[L]$)
and
$\mathcal A^\text{\tiny glob}_{n+\nu}[L]$
($\mathfrak A^\text{\tiny glob}_{n+\nu}[L]$),
the coefficients $\tilde{d}_{n}$ in Eqs.\ (\ref{eq:Ini.Series}) and
(\ref{eq:Sum.DR.Global}) are also different.
In order to obtain a generalization of the resummation procedure,
given by Eq.\ (\ref{eq:FAPT.Sum}), we propose to apply it to the
spectral densities
$\rho_{n+\nu}^\text{\tiny glob}[L_\sigma]$.
This will be done for every $N_f$-fixed integration region
causing its reduction to $\bar{\rho}_{n+\nu}[L_\sigma+\lambda_f;N_f]$.
Subsequently, application of Eq.\ (\ref{eq:FAPT.Sum}) to sum up the
series
$
 \sum\limits_{n\geq1}\bar{\rho}_{n+\nu}[L_\sigma+\lambda_f;N_f]
 \left\langle\!\left\langle{t^{n-1}}\right\rangle\!\right\rangle_{P}
$
will yield
\begin{eqnarray}
\label{eq:Sum.SD.FAPT}
  \sum\limits_{n\geq1}\bar{\rho}_{n+\nu}[L_\sigma+\lambda_f;N_f]
  \left\langle\!\left\langle{t^{n-1}}\right\rangle\!\right\rangle_{P}
&=&
   \left\langle\!\!\!\left\langle{
   \bar{\rho}_{1+\nu}
   \left[L_\sigma+\lambda_f-\frac{t}{\beta_f};N_f\right]}
   \right\rangle\!\!\!\right\rangle_{P_\nu}\, .
\end{eqnarray}
As shown in \cite{BM08}, this procedure generates in the Minkowski
region the following answer for the analytic image of the entire sum:
\begin{eqnarray}
  R_{\nu}^\text{\tiny FAPT}[L]
=
  d_0\,{\mathfrak A}^\text{\tiny glob}_{\nu}[L]
  \!&\!+\!&\!
  d_1\sum\limits_{f=3}^{6}
  \theta\left(L_{f}\!\leq\!L\!<\!L_{f+1}\right)
  \left\langle\!\!\!\left\langle { \bar{\mathfrak A}_{1+\nu}\!
  \Big[L\!+\!\lambda_f\!-\!\frac{t}{\beta_f};f\Big]
                                 }
  \right\rangle\!\!\!\right\rangle_{P_{\nu}}
\nonumber\\
\!&\!+\!&\!
  d_1\sum\limits_{f=3}^{6}
  \theta\left(L_{f}\!\leq\!L\!<\!L_{f+1}\right)\!\!
  \sum\limits_{k=f+1}^{6}\!
  \left\langle\!\left\langle{\Delta_{k}\bar{\mathfrak A}_{1+\nu}[t]}
  \right\rangle\!\right\rangle_{P_{\nu}}~
\label{eq:Glo-MFAPT.sum.R.Glo.456}
\end{eqnarray}
with $ P_{\nu}(t)$ taken from Eq.\ (\ref{eq:P.nu}).

\section{Improved Naive Non-Abelization procedure}
 \label{App:INNA}
Following the analysis of Ref.\ \cite{MS04}, we consider an expansion
of the perturbative coefficients $d_n$ in a power series in
$b_0, b_1,\ldots$, i.e.,
\begin{eqnarray*}
  \tilde{d}_3
=
  b_0^2\,d_3[2,0]
  + b_1\,d_3[0,1]
  + b_0\,d_3[1,0]
  + d_3[0,0]\, ,
\end{eqnarray*}
as opposed to the standard expansion in a power series in $N_f$
(i.e., the number of flavors),
$$
  \tilde{d}_3= N_f^2 \, d_3(2)+N_f^1\, d_3(1)+N_f^0\, d_3(0).
$$
Here, the first argument $n_0$ of the coefficients
$d_n[n_0,n_1,\ldots]$
corresponds to the power of $b_0$,
whereas the second one $n_1$ refers
to the power of $b_1$, etc.
The coefficient $d_n[0,0,\ldots,0]$ represents the ``genuine''
corrections which are associated with the coefficients $b_i$ in the
power $n_i=0$.
If all the arguments of the coefficient
$d_n[\ldots,m,0,\ldots,0]$
to the right of the index $m$ are equal to zero, then, for the sake of
a simplified notation, we will omit these arguments and write instead
$d_n[\ldots,m]$.
Applying this terminology, we obtain the following representation
for the next few coefficients
\begin{eqnarray}
\label{eq:d_3}
  \tilde{d}_3
&=& b_0^2\,\underline{d_3[2]}
     + b_1\,\underline{d_3[0,1]}
     + b_0\,d_3[1]
     + d_3[0]\,,\\
\label{eq:d_4}
  \tilde{d}_4
&=& b_0^3\,\underline{d_4[3]}
     + b_1\,b_0\,\underline{d_4[1,1]}
     + b_2\,\underline{d_4[0,0,1]}
     + b_0^2\,d_4[2]
     + b_1\,d_4[0,1]
     + b_0\,d_4[1]
     + d_4[0]\,, \\
\label{eq:d_5}
  \tilde{d}_5
&=& b_0^4\,\underline{d_5[4]}
     + b_1\,b_0^2\,\underline{d_5[2,1]}
     + b_1^2\,\underline{d_5[0,2]}
     + b_0\,b_2\,\underline{d_5[1,0,1]}
     + b_3\,\underline{d_5[0,0,0,1]}
     + O\!\left(b_0^3\right).
\end{eqnarray}
The same ordering of the $\beta$-function elements applies with
respect to the higher coefficients $d_n$ as well.
Employing the standard NNA approximation, one estimates $\tilde{d}_n$
from the first term in the equations above, namely,
$\tilde{d}_n \simeq b_0^n\underline{d_n[n-1]}$.
We are going to improve this estimate
by including other terms of the same order as $b_0^n$
by means of the relation
$b_i \simeq O(b_0^{(i+1)})$.
[For the reader's convenience, we have underlined the coefficients of
these terms in Eqs.\ (\ref{eq:d_3})--(\ref{eq:d_5}).]
Note that in the $\overline{\text{MS}}$ scheme, the proportionality
coefficients $c_i=b_i/b_0^{(i+1)}$ are
\begin{eqnarray}
 c_1 \approx 0.74\,;~~
 c_2 \approx 0.7\,;~~
 c_3 \approx 1.82\,;~~
 c_4 = \text{unknown}\,.
\end{eqnarray}
All these coefficients are of order unity and their values have
been estimated for $N_f=4$, meaning that there is no reason to
neglect them in Eqs.\ (\ref{eq:d_3})--(\ref{eq:d_5}).
This observation allows one to generalize the large
$b_0$-approximation, by taking into account all terms with the
underlined coefficients in Eqs.\ (\ref{eq:d_3})--(\ref{eq:d_5})
that belong to the same order in $b_0$.
Within this setup, the element $\underline{d_n[n-1]}$ can be easily
obtained from the term $d_n(n-1)$, i.e., via
$\underline{d_n[n-1]}=\left(-\frac{3}{2}\right)^{n-1}d_n(n-1)$.
Moreover, the coefficients $d_n(n-1)$, related to the vector
correlator, and the sum rules pertaining to deep-inelastic
scattering were obtained for any $n$ in
\cite{BroadKa93,BroadKa02}, respectively.

The determination of the remaining underlined elements in Eqs.\
(\ref{eq:d_3})--(\ref{eq:d_5}) is a difficult task that has been
partially carried out in \cite{MS04} for the particular case of the
Adler $D$-function in the representation of Eq.\ (\ref{eq:d_3}).
Strictly speaking, it was found that
$\underline{d_3[0,1]}\approx -0.4\,\underline{d_3[2]}$
and $\underline{d_3[0,1]}\approx 1.2$,
and it was proved that both elements are of order unity.
With this in mind, we propose to use in the expansions of
$D_S$, [cf.\ (\ref{eq:D_S})],
and $D_V$,\, [cf.\ (\ref{eq:D_V})],
the following relation for the considered coefficients
\begin{equation}
  \big|\underline{d_n[n_0,n_1,\ldots]}\big|
=
  \big|\underline{d_n[n-1]}\big| \, .
\label{eq:equal}
\end{equation}
Surprisingly, this rough approximation leads to reasonable results
for the coefficients $d_n$ (especially when compared with those
found with the NNA method), as one can see from the entries
dubbed ``INNA''  in Tables \ref{Tab:r_n.d_k}, \ref{Tab:d_n.Higgs},
\ref{Tab:d_n.Adler}.
As an illustrative example of this procedure, use Eq.\ (\ref{eq:equal})
in Eq.\ (\ref{eq:d_4}) to predict the coefficients
\begin{eqnarray}
  \tilde{d}_4\approx \tilde{d}^\text{INNA}_4
&=&
  d_4[3]\left(b_0^3 +b_1 b_0 \pm b_2\right)+ O(b_0^2) \\
\tilde{d}_5 \approx \tilde{d}^\text{INNA}_5
&=&
  d_5[4]\left(b_0^4 +b_1 b_0^2+b_1^2+ b_0\,b_2 \pm b_3\right)+ O(b_0^3)
\, ,
\end{eqnarray}
with their values being given in Tables \ref{Tab:d_n.Higgs}
and \ref{Tab:d_n.Adler}.
This kind of approximation is based upon the condition (\ref{eq:equal})
and is in line with the underlying assumptions of the original NNA
procedure.

\section{Two-loop results}
 \label{app:Two-loop}
\subsection{Recurrence relations}
The expansion of the $\beta$-function in the two-loop approximation
is given by
\begin{eqnarray}
  \frac{d}{dL}\left(\frac{\alpha_{s}[L]}{4 \pi}\right)
=
  - b_0\left(\frac{\alpha_{s}[L]}{4 \pi}\right)^2
  - b_1\left(\frac{\alpha_{s}[L]}{4 \pi}\right)^3\, ,
\label{eq:betaf}
\end{eqnarray}
where $L=\ln(\mu^2/\Lambda^2)$ and
\begin{eqnarray}
  b_0 = \frac{11}{3}\,C_\text{A} - \frac{4}{3}\,T_\text{R} N_f
\,,\qquad \qquad b_1
=
  \frac{34}{3}\,C_{\text{A}}^{2}
  - \left(4C_\text{F}
  + \frac{20}{3}\,C_\text{A}\right)T_\text{R} N_f
\label{eq:beta0&1}
\end{eqnarray}
with $C_\text{F}=\left(N_\text{c}^{2}-1\right)/2N_\text{c}=4/3$,
$C_\text{A}=N_\text{c}=3$, $T_\text{R}=1/2$, and $N_f$ denoting
the number of active flavors.
Then, the corresponding two-loop equation for the coupling
$a=b_0\,\alpha/(4\pi)$ reads
\begin{eqnarray}
  \frac{d a_{(2)}[L]}{dL}
= - a_{(2)}^2[L]\left(1 + c_1\,a_{(2)}[L]\right)
  \quad \text{with}~c_1\equiv\frac{b_1}{b_0^2}\, .
\label{eq:beta.new}
\end{eqnarray}
Still higher beta-function coefficients, e.g., $b_2$, $b_3$,
can be found in \cite{vRVL97,MS04}.
This equation immediately generates the following recurrence relation
\begin{subequations}
\begin{eqnarray}
  -\frac{1}{n}\,
   \frac{d a_{(2)}^n[L]}{dL}
=
   a_{(2)}^{n+1}[L]
  +c_1\,a_{(2)}^{n+2}[L]
\label{eq:rec.rel.n.n+2.PT}
\end{eqnarray}
for consecutive powers of the coupling constant.
Due to its linearity, this relation remains valid also for the
analytic images of the coupling's powers:
\begin{eqnarray}
  -\frac{1}{n}\,
   \frac{d}{dL}\,
   \mathcal F_n[L]
=
   \mathcal F_{n+1}[L]
   + c_1\,\mathcal F_{n+2}[L]\, ,
\label{eq:rec.rel.n.n+2.APT}
\end{eqnarray}
where, as it has already been used in Sec.\ \ref{sec:APT.Resummation},
${\mathcal F}[L]$ denotes one of the analytic quantities
$\mathcal A^{(2)}[L],~\mathfrak A^{(2)}[L]$, or $\rho^{(2)}[L]$.
Quite analogously, we obtain the following generalization of this
relation, pertaining to fractional coupling-constant indices, viz.,
\begin{eqnarray}
  -\frac{1}{n+\nu}\, \frac{d}{dL}\, \mathcal F_{n+\nu}[L]
=
  \mathcal F_{n+1+\nu}[L]
  + c_1\,\mathcal F_{n+2+\nu}[L]\, .
\label{eq:rec.rel.n.n+2.FAPT}
\end{eqnarray}
\end{subequations}

In the two-loop approximation we have a different evolution
for the running mass,
which reads
\begin{subequations}
\label{eq:mass.Ev.2L}
\begin{eqnarray}
  \bar{m}_{b}^2(Q^2)
&=&
  \hat{m}_{(2)}^2\,
  \left[\alpha_{s}^{(2)}(Q^2)\right]^{\nu_{0}}
  \left[1 + \frac{c_1\,b_0}{4\,\pi}\,
  \alpha_{s}^{(2)}(Q^2)\right]^{\nu_1}\\
&=&
  \hat{m}_{(2)}^2\,
  \left[\frac{4\,\pi}{b_0}\,a_{(2)}(Q^2)\right]^{\nu_{0}}
  \left[1 + c_1\,a_{(2)}(Q^2)\right]^{\nu_1} \, ,
\end{eqnarray}
\end{subequations}
where
\begin{eqnarray}
 \nu_1 = 2 \left(\frac{\gamma_1}{b_1}
                -\frac{\gamma_0}{b_0}
           \right)
\end{eqnarray}
and $\hat{m}_{(2)}$
is the renormalization-group-invariant mass.

\subsection{Resummation in FAPT for fixed $N_f$}
Consider here the following power series with $\nu\geq0$:
\begin{eqnarray}
  \mathcal W_\nu[L;t;\mathcal F]
\equiv
  \sum\limits_{n\geq1}
  t^{n-1}
  \mathcal F_{n+\nu}[L]
\label{eq:series.W.FAPT}
\end{eqnarray}
noting that for $\nu=0$ we would obtain the corresponding two-loop
APT expression.
Multiplying both sides of Eq.\ (\ref{eq:rec.rel.n.n+2.FAPT})
with $t^{n+\nu}$ and summing over $n$ from $n=1$ to $\infty$,
we arrive at
\begin{eqnarray*}
  -\hat{p}\!\int\limits_{0}^{t}\!t'^\nu\,
  \mathcal W_{\nu}[L;t';\mathcal F] \,dt'
=
  t^\nu\,
  \left[
        \left(1+\frac{c_1}{t}\right)\,
        \left(\mathcal W_{\nu}[L;t;\mathcal F]
             -\mathcal F_{1+\nu}[L]\right)
        - c_1\,\mathcal F_{2+\nu}[L]
  \right]\, ,
\end{eqnarray*}
where $\hat{p}\equiv d/dL$.
Differentiating this equation with respect to $t$, we obtain
\begin{eqnarray}
  \frac{d}{dt}
   \left\{
          t^\nu
          \left[\left(1+\frac{c_1}{t}\right)\,
                \left(\mathcal W_{\nu}[L;t;\mathcal F]
                    - \mathcal F_{1+\nu}[L]
                \right)
             - c_1\,\mathcal F_{2+\nu}[L]
         \right]
   \right\}
           + \hat{p}\,t^\nu\,\mathcal W_{\nu}[L;t;\mathcal F]
=
  0
\label{eq:difur.W.FAPT}
\end{eqnarray}
with the initial condition
$\mathcal W_{\nu}[L;0;\mathcal F]=\mathcal F_{1+\nu}[L]$.
In order to solve this equation, we consider the following function
\begin{subequations}
\begin{eqnarray}
  \mathcal X_{\nu}[L;t]
&\equiv&
  \left(1+\frac{c_1}{t}\right)
  \sum\limits_{n\geq2}
                      t^{n+\nu-1}
                      \mathcal F_{n+\nu}[L]
  - c_1\,t^{\nu}\mathcal F_{2+\nu}[L]\, ,
\label{eq:series.X.FAPT}\\
  \mathcal X_{\nu}[L;0]
&=&
  0\, .
\label{eq:X[L;0].FAPT}
\end{eqnarray}
\end{subequations}
Our initial series $\mathcal W[L;t;\mathcal F]$ is related to this
function by the evident relation
\begin{eqnarray}
  \mathcal W_{\nu}[L;t;\mathcal F]
&=&
  \mathcal F_{1+\nu}[L]
  + \frac{t^{1-\nu}}{c_1+t}\,\mathcal X_\nu[L;t]
  + \frac{c_1\,t}{c_1+t}\,\mathcal F_{2+\nu}[L]
 \, .
\label{eq:X.W.FAPT}
\end{eqnarray}
Hence, we find
\begin{eqnarray*}
  \frac{d}{dt}\,\mathcal X_\nu[L;t]
  + \hat{p}\,\frac{t}{c_1+t}\,\mathcal X_\nu[L;t]
&=& - \hat{p}\,t^\nu\,\mathcal F_{1+\nu}[L]
    - \hat{p}\,\frac{c_1\,t^{1+\nu}}{c_1+t}\,\mathcal F_{2+\nu}[L] \, .
\end{eqnarray*}
Finally, using the substitution
\begin{eqnarray}
  \mathcal X_\nu[L;t]
=
  e^{-\hat{p}\tau(t)} X_\nu[L;t] \, ,
\end{eqnarray}
we get
\begin{eqnarray*}
  \frac{d X_\nu[L;t]}{dt}
=
  -\hat{p}\,
  e^{\hat{p}\tau(t)}
  \left(
        t^\nu\,\mathcal F_{1+\nu}[L]
        + \frac{c_1\,t^{1+\nu}}{c_1+t}\,\mathcal F_{2+\nu}[L]
  \right)\, .
\end{eqnarray*}
That implies the relations
\begin{eqnarray*}
  X_\nu[L;t]
&=&
  - t^{1+\nu}\!
  \int\limits_{0}^{1}\!z^\nu\,dz\,
  \left[
        \frac{d\mathcal F_{1+\nu}[L+\tau(t\,z)]}{dL}
        + \frac{c_1\,t\,z}{c_1+t\,z}\,
        \frac{d\mathcal F_{2+\nu}[L+\tau(t\,z)]}{dL}
  \right]
  \,;\\
  t^{-\nu}\mathcal X_\nu[L;t]
&=&
  - t\!\int\limits_{0}^{1}\!z^\nu dz\,
  \left[
        \frac{d\mathcal F_{1+\nu}[L+\tau(t\,z)-\tau(t)]}{dL}
        + \frac{c_1\,t\,z}{c_1+t\,z}\,
        \frac{d\mathcal F_{2+\nu}[L+\tau(t\,z)-\tau(t)]}{dL}
  \right]\, ,
\end{eqnarray*}
where we took into account that $\hat{p}=d/dL$ and where we used
the initial condition (\ref{eq:X[L;0].FAPT}).
Making use of this solution in Eq.\ (\ref{eq:X.W.FAPT}), we obtain
\begin{eqnarray*}
  \mathcal W_{\nu}[L;t;\mathcal F]
&=&
  \mathcal F_{1+\nu}[L]
  + \frac{t}{c_1+t}\,
  \Big\{
        c_1\,\mathcal F_{2+\nu}[L]
        - t\!\int\limits_{0}^{1}\!z^\nu dz\,
        \Big[\frac{d\mathcal F_{1+\nu}[L+\tau(t\,z)-\tau(t)]}{dL}
~~~~~~~~~~~\nonumber\\
& &~~~~~~~~~~~~~~~~~~~~~~~~~~~~~~~~~~~~~~
        + \frac{c_1\,t\,z}{c_1+t\,z}\,
        \frac{d\mathcal F_{2+\nu}[L+\tau(t\,z)-\tau(t)]}{dL}
        \Big]
  \Big\}\, .
\end{eqnarray*}
Note that this equation can be recast in the equivalent form
\begin{eqnarray}
  \mathcal W_{\nu}[L;t;\mathcal F]
&=&
  \mathcal F_{1+\nu}[L]
 - \frac{t^2}{c_1+t}\!\int\limits_{0}^{1}\!z^\nu dz\,
               \frac{d\mathcal F_{1+\nu}[L+\tau(t\,z)-\tau(t)]}{dL}
    \nonumber\\
&+& \frac{c_1\,t}{c_1+t}\,
     \Big\{\mathcal F_{2+\nu}[L-\tau(t)]\,\delta_{0,\nu}
         + \nu\!\int\limits_{0}^{1}\!z^{\nu-1}\,
            \mathcal F_{2+\nu}[L+\tau(t\,z)-\tau(t)]\,dz
     \Big\}\, , ~~~
\label{eq:W[L;x].exact.FAPT}
\end{eqnarray}
with $\delta_{0,\nu}$ being a Kronecker delta symbol.
The reason for this recasting is that it is more appropriate for
realizing the limit $c_1\to 0$ and obtain this way the one-loop
expressions, given by Eqs.\ (\ref{eq:FAPT.Sum})--(\ref{eq:P.nu}).
We observe that the analytic sum of the initial power series in $t$
can be represented by means of the analytic couplings
$\mathcal F_{1+\nu}[L]$ and $\mathcal F_{2+\nu}[L]$.
Setting $\nu=0$, we obtain the corresponding two-loop APT expression
\begin{eqnarray}
  \mathcal W_{0}[L;t;\mathcal F]
&=& \mathcal F_1[L]
  - \frac{t}{c_1+t}\!
  \int\limits_{0}^{t}\!
  \frac{d\mathcal F_1[L+\tau(t')-\tau(t)]}{dL}\,dt'
  + \frac{c_1\,t}{c_1+t}\,
  \mathcal F_2[L-\tau(t)]\, . ~~
\label{eq:W[L;x].exact.APT}
\end{eqnarray}

Hence, the sum expressed via equations like (\ref{eq:FAPT.Sum}) can be
recast in convolution form to read\footnote{
Here we set $\mathcal S=D^\text{\tiny FAPT}$ or $R^\text{\tiny FAPT}$
depending on the specific choice of
$\mathcal F=\mathcal A^\text{\tiny glob}$ or
$\mathcal F=\mathfrak A^\text{\tiny glob}$.}
\begin{eqnarray}
  \mathcal S_{\nu}[L]
=
  d_0\mathcal F_\nu[L]
  + d_1\,
         \left\langle\!\!\!\left\langle
         \mathcal F_{1+\nu}[L]
         + \frac{c_1\,t}{c_1+t}\,\mathcal F_{2+\nu}[L-\tau(t)]\,
         \delta_{0,\nu}
         \right\rangle\!\!\!\right\rangle_{\!\!\!P}
 ~~~~~~~~~~~~~~~~~~~\\
 ~+ \left\langle\!\!\!\left\langle
     \frac{-\,d_1\,t^2}{c_1+t}\,
      \int\limits_{0}^{1}\!z^\nu dz\,
       \left\{\frac{d\mathcal F_{1+\nu}[L+\tau(t\,z)-\tau(t)]}{dL}
            + \frac{c_1\,\nu}{t\,z}\,
               \mathcal F_{2+\nu}[L+\tau(t\,z)-\tau(t)]
       \right\}
    \right\rangle\!\!\!\right\rangle_{\!\!\!P} \, .
\nonumber
\label{eq:S[L;x].exact.FAPT}
\end{eqnarray}
On the other hand, for the APT case with $\nu=0$, expressions like
(\ref{eq:Ini.Series}) can be rewritten as
\begin{eqnarray}
\label{eq:S[L;x].exact.APT}
  \mathcal S[L]
= d_0
  + d_1\,
  \left\langle\!\!\!
              \left\langle
  \mathcal F_1[L]
  - \frac{t}{c_1+t}\!
  \int\limits_{0}^{t}\!
       \frac{d\mathcal F_1[L+\tau(t')-\tau(t)]}{dL}\,dt'
       + \frac{c_1\,t}{c_1+t}\,
       \mathcal F_2[L-\tau(t)]
               \right\rangle\!\!\!
  \right\rangle_{\!\!\!P}\,.~~~~~
\end{eqnarray}

Quite analogously it is possible to resum also the following expression
\begin{eqnarray}
\label{eq:Sum.2L.Evo}
  \mathcal E_{\nu;\nu_1}[L]
= d_0\,\mathcal B_{\nu;\nu_1}[L]
  + d_1\sum_{n=0}^{\infty}
  \frac{\left\langle\!\left\langle{(-t)^{n}}\right\rangle\!
  \right\rangle_{P}}
  {n!}\,
  \mathcal B_{n+1+\nu;\nu_1}[L]\, ,~
\end{eqnarray}
in which
$
 \mathcal B_{\nu;\nu_1}[L]
=
 \textbf{A}_{E,M}\left[a_{(2)}^{\nu}[L]
 \left(1+c_1\,a_{(2)}\right)^{\nu_1}[L]\right]
$
is the analytic image of the two-loop evolution factor
(\ref{eq:mass.Ev.2L}).
Then, the FAPT result for the resummation of this series is
\begin{eqnarray}
  \mathcal E_{\nu;\nu_1}[L]
\!&\!=\!&\!
  d_0\mathcal B_{\nu;\nu_1}[L] + d_1\,
  \left\langle\!\!\!\left\langle
    \mathcal B_{1+\nu;\nu_1}[L]
      + \frac{c_1\,t}{c_1+t}\,\mathcal B_{2+\nu;\nu_1}[L]
       \right\rangle\!\!\!\right\rangle_{\!\!\!P}\nonumber\\
\!&\!-\!&\!
  \left\langle\!\!\!\left\langle
    \frac{d_1\,t^2}{(c_1+t)^{1-\nu_1}}\,
     \int\limits_{0}^{1}\!dz\,
      \frac{z^{\nu+\nu_1}}{(c_1+t\,z)^{\nu_1}}
       \frac{d\mathcal B_{1+\nu;\nu_1}[L+\tau(t\,z)-\tau(t)]}{dL}
  \right\rangle\!\!\!\right\rangle_{\!\!\!P}
~~~\nonumber\\
 \!&\!-\!&\!
  \left\langle\!\!\!\left\langle
    \frac{d_1\,c_1\,t}{(c_1+t)^{1-\nu_1}}\,
     \int\limits_{0}^{1}\!dz\,
      \frac{z^{\nu+\nu_1}}{(c_1+t\,z)^{\nu_1}}
       \frac{t^2\,z}{c_1+t\,z}\,
        \frac{d\mathcal B_{2+\nu;\nu_1}[L+\tau(t\,z)-\tau(t)]}{dL}
  \right\rangle\!\!\!\right\rangle_{\!\!\!P}
\nonumber\\
 \!&\!+\!&\!
   \left\langle\!\!\!\left\langle
    \frac{d_1\,c_1\,t^2}{(c_1+t)^{1-\nu_1}}\,
     \int\limits_{0}^{1}\!dz\,
      \frac{\nu_1\,z^{\nu+\nu_1}}{(c_1+t\,z)^{1+\nu_1}}\,
       \mathcal B_{2+\nu;\nu_1}[L+\tau(t\,z)-\tau(t)]
   \right\rangle\!\!\!\right\rangle_{\!\!\!P} \,.~~~
 \label{eq:Resum.2L.Evo}
\end{eqnarray}
Using the exact derivative expression
\begin{eqnarray*}
  \frac{d\mathcal B_{2+\nu;\nu_1}[L+\tau(t\,z)-\tau(t)]}{dz}
=
  \left(\frac{t^2\,z}{c_1+t\,z}\right)
  \frac{d\mathcal B_{2+\nu;\nu_1}[L+\tau(t\,z)-\tau(t)]}{dL}
\end{eqnarray*}
and integrating by parts we can rewrite
(\ref{eq:Resum.2L.Evo}) in a more convenient way to read
\begin{eqnarray}
  \mathcal E_{\nu;\nu_1}[L]
&\!=\!&
  d_0\,\mathcal B_{\nu;\nu_1}[L]
  + d_1\left\langle\!\!\!\left\langle
           \mathcal B_{1+\nu;\nu_1}[L]
         + \delta_{0,{\nu+\nu_1}}\,
           t\,\left[\frac{c_1}{c_1+t}\right]^{1-\nu_1}
           \mathcal B_{2+\nu;\nu_1}[L-\tau(t)]
       \right\rangle\!\!\!\right\rangle_{\!\!\!P}\nonumber\\
\!&\!-\!&\!
  \left\langle\!\!\!\left\langle
    \frac{d_1\,t^2}{(c_1+t)^{1-\nu_1}}\,
     \int\limits_{0}^{1}\!dz\,
      \frac{z^{\nu+\nu_1}}{(c_1+t\,z)^{\nu_1}}
       \frac{d\mathcal B_{1+\nu;\nu_1}[L+\tau(t\,z)-\tau(t)]}{dL}
  \right\rangle\!\!\!\right\rangle_{\!\!\!P}~~~
\nonumber\\
 \!&\!+\!&\!
  \left\langle\!\!\!\left\langle
    \frac{d_1\,c_1\,t}{(c_1+t)^{1-\nu_1}}\,
     \int\limits_{0}^{1}\!dz\,
      \frac{(\nu+\nu_1)\,z^{\nu+\nu_1-1}}{(c_1+t\,z)^{\nu_1}}\,
       \mathcal B_{2+\nu;\nu_1}[L+\tau(t\,z)-\tau(t)]
  \right\rangle\!\!\!\right\rangle_{\!\!\!P} \, .
\label{eq:Resum.2L.Evo.imp}
\end{eqnarray}

\subsection{Resummation in global FAPT}
The formalism developed in the text and the previous appendices can be
applied to the global case using spectral densities, which correspond
to a fixed number of active flavors $N_f$, and are defined in the
integration intervals
$L_{f}<L_{\sigma}\leq L_{f+1}$.
Indeed, employing
$\mathcal F_\nu[L]=\rho_\nu[L]$ and $L=L_\sigma$,
we can first perform the resummation before carrying out the spectral
integration over $L_\sigma$.

Let us study these operations in some more detail within the
Euclidean FAPT.
In that case, the global spectral density
$\rho^\text{\tiny glob}_{n+\nu}(\sigma)$
has the form specified in Eq.\ (\ref{eq:rho_n.Glo}).
It can be rewritten in an equivalent way to read
\begin{eqnarray}
  \rho^\text{\tiny glob}_{n+\nu}[L_\sigma]
= \sum\limits_{f=3}^{6}
  \theta\left[L_{f}<L_\sigma\leq L_{f+1}\right]\,
  \bar{\rho}_{n+\nu}[L_\sigma+\lambda_f;N_f]\, ,
\label{eq:rho_n.Glo_equiv}
\end{eqnarray}
where we used $L_3=-\infty$ and $L_7=+\infty$.
Then, we have for the sum of the global power series (in $t$) the
following expression
\begin{eqnarray}
  \mathcal W^\text{\tiny FAPT}_\nu[L;t]
\equiv
  \sum\limits_{n\geq1}
  t^{n-1}
  \mathcal A^\text{\tiny glob}_{n+\nu}[L]
=
  \sum\limits_{f=3}^{6}
  \int_{L_f}^{L_{f+1}}\!
  \frac{\rho_{\nu}^{\text{sum}}\left[L_{\sigma}+\lambda_f;N_f;t\right]\,
  dL_\sigma}
  {1+e^{L-L_{\sigma}}}
\label{eq:series.W.FAPT.glob}
\end{eqnarray}
in which the following abbreviation was used:
\begin{eqnarray}
\label{eq:rho.nu.t.sum}
  \rho_{\nu}^{\text{sum}}\left[L;N_f;t\right]
&\equiv&
  \sum\limits_{n\geq1}
  t^{n-1}
  \bar{\rho}_{n+\nu}[L;N_f]
= \left(\frac{1}{\beta_f}\right)^{1+\nu}
  \sum\limits_{n\geq1}
  \left[\frac{t}{\beta_f}\right]^{n-1}
  \rho_{n+\nu}[L]\,.
\end{eqnarray}
We see that this series has the form of Eq.\ (\ref{eq:series.W.FAPT})
and is equal to
\begin{eqnarray}
\label{eq:rho.nu.t.resum}
  \rho_{\nu}^{\text{sum}}\left[L;N_f;t\right]
&=&
  \left(\frac{1}{\beta_f}\right)^{1+\nu}\,
  \mathcal W_\nu\left[L;\frac{t}{\beta_f};\rho^{(2)}\right]\, ,
\end{eqnarray}
where the last argument
$\rho^{(2)}$ of $\mathcal W_\nu$
means that everywhere in Eq.\ (\ref{eq:W[L;x].exact.FAPT})
one should substitute $\mathcal F[L]\to \rho^{(2)}[L]$.
Hence, the initial sum
\begin{eqnarray}
 D^\text{\tiny FAPT}_\nu[L]
  \equiv
  d_0\mathcal A^\text{\tiny glob}_\nu[L]
  + d_1\,
  \sum\limits_{n\geq1}
  \left\langle\!\left\langle
  {t^{n-1}\mathcal A^\text{\tiny glob}_{n+\nu}[L]}
  \right\rangle\!\right\rangle_{\!P}
\label{eq:series.S.FAPT.glob}
\end{eqnarray}
can be resummed in the following final form
\begin{subequations}
\begin{eqnarray}
 D^\text{\tiny FAPT}_\nu[L]
 &=& d_0\mathcal A^\text{\tiny glob}_\nu[L]
  + d_1\,\sum\limits_{f=3}^{6}
  \int_{L_f}^{L_{f+1}}\!
  \frac{\rho_{\nu}^{\text{sum}}\left[L_{\sigma}+\lambda_f;N_f\right]\,
  dL_\sigma}
  {1+e^{L-L_{\sigma}}}
 \label{eq:S[L;x].exact.FAPT.glob}\\
  \rho_{\nu}^{\text{sum}}\left[L;N_f\right]
 &\equiv&
  \left(\frac{1}{\beta_f}\right)^{1+\nu}\,
  \left\langle\!\!\!\left\langle
  \mathcal W_\nu\left[L;\frac{t}{\beta_f};\rho^{(2)}\right]
  \right\rangle\!\!\!\right\rangle_{\!\!\!P} \, .
 \label{eq:rho.nu.resum}
\end{eqnarray}
\end{subequations}

\section{Pole and RG-invariant masses of the bottom quark}
 \label{app:masses.b-quark}
We start our considerations by writing down the relation
\cite{ChSt99,ChSt00} between the pole mass of the $b$ quark, $m_b$,
on one hand, and the value of the running mass at the scale
$\mu_{*}=\overline{m}_b(\mu_{*}^2)$,
which we call $\overline{m}_b(\overline{m}_b^2)$,
on the other hand:
\begin{eqnarray}
  \label{eq:mb-pole.mb(mb)}
  \left[\frac{\overline{m}_b(\overline{m}_b^2)}
             {m_b}
  \right]^2
=
  1-\frac{8}{3}\,\frac{\alpha_s\left(m_b^{2}\right)}{\pi}
  -13.234\,\left[\frac{\alpha_s\left(m_b^{2}\right)}{\pi}\right]^2
  \,.
\end{eqnarray}
Note that some authors prefer another version of this relation
in which the pole mass $m_b$ is used as argument in the numerator of
the left-hand side: $\overline{m}_b(m_b^2)$---see
\cite{ChSt99,ChSt00,GBGS90,KK08}.\footnote{
We wish to thank A.~Kataev for attracting our attention to this point.}

The QCD scales are fixed via the normalization of the strong
coupling in the corresponding one- or two-loop approximation at the
$Z$-pole, employing the condition
$R_{e^+e^-\to\text{hadrons}}(\alpha_s(m_Z^2))=1.039$,
suggested in~\cite{BCK08},
where
\begin{eqnarray}
\label{eq:R.e+e-.to.hadrons}
  R_{e^+e^-\to\text{hadrons}}(\alpha_s)
&=&
  1 + \frac{\alpha_s}{\pi}
    + 1.4097\,\left(\frac{\alpha_s}{\pi}\right)^2\, .
\end{eqnarray}

\begin{table}[h]
 \caption{The values of the QCD scale parameters $\Lambda_{f}$
  and $b$-quark masses pertaining to \cite{KuSt01}.
  \label{Tab:m_b.Lambda}}
 \centerline{%
  \begin{tabular}{|c|c|c|c|c|c|}
   \hline
    ~Loop order~   &~$\Lambda_{3}$~[MeV]~
                         &~$\Lambda_{5}$~[MeV]~
                                &~$\overline{m}_b(\overline{m}_b^2)$~[GeV]
                                        &~${m_b}$~[GeV]~
                                               &~$\hat{m}_b$~[GeV]
                                               \\ \hline
    1-loop 
                   & 184 & 115  & 4.21  & 4.69 & 8.22 \\
    2-loop 
                   & 346 & 195  & 4.19  & 4.84 & 7.89 \\
\hline
  \end{tabular}
}
\end{table}
\begin{table}[h]
 \caption{The values of the $b$-quark masses employing the estimates
  of the Penin--Steinhauser paper \cite{PeSt02}.
  \label{Tab:m_b.Lambda.Penin}}
 \centerline{%
  \begin{tabular}{|c|c|c|c|c|c|}
   \hline
    ~Loop order~   &~$\Lambda_{3}$~[MeV]~
                         &~$\Lambda_{5}$~[MeV]~
                                &~$\overline{m}_b(\overline{m}_b^2)$~[GeV]
                                        &~${m_b}$~[GeV]~
                                               &~$\hat{m}_b$~[GeV]
                                               \\ \hline
    1-loop 
                   & 184 & 115  & 4.35  & 4.84 & 8.53 \\
    2-loop 
                   & 346 & 195  & 4.34  & 5.00 & 8.22 \\
 \hline
  \end{tabular}}
\end{table}

Tables \ref{Tab:m_b.Lambda} and \ref{Tab:m_b.Lambda.Penin}
show the results for $\hat{m}_b$ obtained in this work.
In parallel, we present results for the pole mass $m_b$,
using as input the estimates for
$\overline{m}_b(\overline{m}_b^2)$
derived by K\"{u}hn and Steinhauser in Ref.\
\cite{KuSt01}---Table \ref{Tab:m_b.Lambda}.
In Table \ref{Tab:m_b.Lambda.Penin} we present analogous results,
determining $\hat{m}_b$ from the estimates for
$\overline{m}_b(\overline{m}_b^2)$ derived by Penin and Steinhauser
\cite{PeSt02}.
For the sake of completeness, we also quote here the recent estimates
$\overline{m}_b(\overline{m}_b^2)=4.163\pm0.016$~GeV \cite{Chet09}
and $\overline{m}_b(\overline{m}_b^2)=4.164\pm0.025$~GeV \cite{Kuhn08}.

\end{appendix}


\end{document}